\documentclass[useAMS,usenatbib]{mn2e}
\usepackage{graphicx}
\usepackage{rotating}
\usepackage[section] {placeins}
\usepackage{amsmath, amssymb}
\usepackage{color}
\usepackage{multirow}
\usepackage{makeidx}
\usepackage{xspace}
\usepackage{float}
\usepackage{subfig}

\newcommand{\degree}{$^{\circ}$}

\newcommand{\kms}[0]{\ensuremath{km~s$^{-1}$}\xspace}

\def\farcs{\hbox{$.\!\!^{\prime\prime}$}}
\def\fp{\hbox{$.\!\!^{\reset@font\scriptscriptstyle\r@mn{p}}$}}
\def\arcmin{\hbox{$^\prime$}\xspace}
\def\arcsec{\hbox{$^{\prime\prime}$}\xspace}

\def\pd{\partial}

\long\def\crap#1{}

\def\kms{km~s$^{-1}$\xspace}

\def\log{\textrm{log}}

\def\n#1{\textrm{\tiny#1}}
\def\<{\langle }
\def\>{\rangle }
\def\mue{\ensuremath{\<\mu\>_{e}}\xspace}

\def\KS{Kolmogorov-Smirnov\xspace}

\def\A1689{Abell~1689}

\def\KS{Kolmogorov-Smirnov\xspace}

\def\A3D{ATLAS$^\textrm{3D}$\xspace}
\def\SAURON{{\sc SAURON}\xspace}
\def\ppxf{{pPXF}\xspace}
\def\LACOSMIC{{\sc LACosmic}}

\def\fov{field of view\xspace}

\def\lam{$\lambda$\xspace}

\def\eps{$\epsilon$\xspace}

\def\Re{$R_{e}$\xspace}
\def\mue{$\langle\mu_{e}\rangle$\xspace}
\def\V{$V$\xspace}
\def\s{$\sigma$\xspace}

\def\ngal{27\xspace}

\def\sig3{$\Sigma_{3}$\xspace}
\def\rprime{r$^\prime$\xspace}
\def\gprime{g$^\prime$\xspace}

\def\MK{M$_\textrm{K}$\xspace}
\def\grcol{g$^\prime-$r$^\prime$\xspace}
\def\sextractor{{\sc sextractor}\xspace}

\def\sampradius{15\arcmin\xspace}
\def\aprprimelimit{$r^\prime=16$~mag\xspace}
\def\absKlimit{$M_K=-21.5$~mag\xspace}
\def\lameps{$\lambda-\epsilon$\xspace}

\def\velrange{$\pm7000$~km s$^{-1}$\xspace}
\def\nedspeccomplete{\rprime=17.0\xspace}
\def\fsr{$f_{SR}$\xspace}
\def\SRfrac{$0.15\pm{0.06}$\xspace} 
\def\SRpc{$15 \pm{6} \%$\xspace}

\def\SRdiv{$0.31\sqrt{\epsilon}$\xspace}
\def\NSR{$4\pm_{1.6}^{1.7}$\xspace}

\def\pepsdist{0.998}
\def\pmagdist{0.757}

\def\SWIFT{{\sc SWIFT}}

\title[A SWIFT IFS study of the Coma Cluster]{Fast and Slow Rotators in the Densest Environments: a SWIFT IFS study of the Coma Cluster}
\author[R. Houghton et al.]{R. C. W. Houghton$^{1}$\thanks{Email: rcwh@astro.ox.ac.uk}, 
Roger L. Davies$^{1}$, 
F. D'Eugenio$^{1}$, 
N. Scott$^{2}$, 
N. Thatte$^{1}$, 
\newauthor
F. Clarke$^{1}$, 
M. Tecza$^{1}$, 
G. S. Salter$^{3}$, 
L. M. R. Fogarty$^{4}$, 
T. Goodsall$^{1}$\\
$^{1}$Physics Department, University of Oxford, Denys Wilkinson Building, Keble Road, Oxford, OX1 3RH, UK\\
$^{2}$Centre for Astrophysics \& Supercomputing, Swinburne University of Technology, PO Box 218, Hawthorn, VIC 3122, Australia\\
$^{3}$School of Physics, The University of New South Wales, Sydney, NSW 2052, Australia\\
$^{4}$Sydney Institute for Astronomy (SIfA), School of Physics, The University of Sydney, NSW 2006, Australia\\}
\begin{document}

\date{}

\pagerange{\pageref{firstpage}--\pageref{lastpage}} \pubyear{2012}

\maketitle

\label{firstpage}

\begin{abstract}

We present integral-field spectroscopy of \ngal galaxies in the Coma cluster observed with the Oxford \SWIFT\ spectrograph, exploring the kinematic morphology-density relationship in a cluster environment richer and denser than any in the \A3D survey. Our new data enables comparison of the kinematic morphology relation in three very different clusters (Virgo, Coma and Abell~1689) as well as to the field/group environment. 
The Coma sample was selected to match the parent luminosity and ellipticity distributions of the early type population within a radius \sampradius (0.43 Mpc) of the cluster centre, and is limited to \aprprimelimit (equivalent to \absKlimit), sampling one third of that population. From analysis of the \lameps diagram, we find \SRpc of early type galaxies are slow rotators; this is identical to the fraction found in the field and the average fraction in the Virgo cluster, based on the \A3D data. It is also identical to the average fraction found recently in Abell~1689 by \citeauthor{DEugenio2013}. Thus it appears that the \emph{average} slow rotator fraction of early type galaxies remains remarkably constant across many different environments, spanning five orders of magnitude in galaxy number density. However, within each cluster the slow rotators are generally found in regions of higher projected density, possibly as a result of mass segregation by dynamical friction. These results provide firm constraints on the mechanisms that produce early type galaxies: they must maintain a fixed ratio between the number of fast rotators and slow rotators while also allowing the total early-type fraction to increase in clusters relative to the field. A complete survey of Coma, sampling hundreds rather than tens of galaxies, could probe a more representative volume of Coma and provide significantly stronger constraints, particularly on how the slow rotator fraction varies at larger radii.

\end{abstract}


\section{Introduction}
\label{sec:introduction}

Studying the mechanisms that give rise to different galaxy morphologies is central to understanding galaxy formation and evolution. Although considerable progress has been made in reproducing the global characteristics of both late type galaxies (LTGs) and early type galaxies (ETGs), the picture is far from complete.  
A source of confusion for such studies is the fact that visual morphologies do not always map simply to physical characteristics, particularly for the ETGs. The lenticular and elliptical division is not only difficult to measure quantitatively (for it is most commonly made by eye, which is difficult to link to models), but is now known to be degenerate with regard to certain intrinsic properties of the galaxies: for example, the \SAURON and \A3D surveys \citep{SAURONI,ATLAS3DI} found the velocity maps of many ellipticals to be indistinguishable from those of S0s. Furthermore, the same authors identified a clear division in the properties of the velocity maps: most exhibited rapid disk-like rotation, while others showed little or no rotation, leading to the classifications \emph{fast rotator} (FR) and \emph{slow rotator} (SR). These classifications \citep[and sub-classes, see][]{SAURONXII,ATLAS3DII} are based on quantitative analysis of the morphology of velocity maps.

Combining \lam (a proxy for the specific angular momentum) with ellipticity (\eps), the \lameps diagram takes on a similar role to the $V/\sigma-\epsilon$ diagram \citep{Binney78,Davies1983,Binney2005} and can be used to relate the FRs to a family of oblate axisymmetric spheroids \citep{SAURONX,ATLAS3DIII}. The anisotropy of these oblate spheroids is consistent with flattening along the axis perpendicular to the plane of rotation ($z$): flatter galaxies are more anisotropic. Projection effects can then explain the region of the \lameps diagram occupied by the FRs (by assuming a Gaussian distribution of intrinsic ellipticities together with an upper limit in the anisotropy). However, the SRs are not represented by such models. They are an entirely different class of object and may be mildly triaxial \citep{ATLAS3DIII}. 

The prevalence of ETGs in denser, crowded environments (such as galaxy clusters) has long been known \citep{Oemler74,DavisGeller76} with \citet{Dressler1980} parameterising observational evidence in the morphology-density ($T-\Sigma$) relation. However, environment is not adequately described by a single parameter, such as projected density. As discussed by \citet{Muldrew2012}, there are many different environments, and many different measures of environment.  It is also important to realise that there are environments within environments: a massive, dense galaxy cluster may contain under dense regions. This latter case is of particular interest given the results presented later and we find it useful to define the \emph{global host environment} (GHE, such as field, group or cluster) and the \emph{local point environment} (LPE, such as the projected density at the position of a particular galaxy). GHE indicates the scale of largest (host) dark matter halo in the system, while LPE reflects the environment at the precise location of the galaxy in question. Using these definitions, galaxies in the Coma cluster have a cluster GHE, but could have very different LPEs.  Similarly, the $T-\Sigma$ relation tells us how the relative fractions of ellipticals, S0s and Spirals changes with LPE; we remain ignorant about changes with GHE unless we assume a link between LPE and GHE (clusters are more likely to harbour denser LPEs). 
\citet{ATLAS3DVII} revisited the $T-\Sigma$ problem in light of the new SR and FR classification scheme. The updated kinematic morphology-density relation ($kT-\Sigma$) has similar properties to the original: the number of spirals decreases as the number of ETGs increases at higher densities. However the overall fractions of FRs and SRs do not behave in the same manner. While the overall fraction of FRs increases in response to the decrease in the overall fraction of spirals, the overall fraction of SRs increases much more slowly. In fact, when we consider just the fraction of SRs in the ETG population (which we hereafter refer to as the SR fraction or \fsr), it is independent of the LPE density except for a sudden increase at the highest densities from around 15\% to 25\% \citep[c.f. Fig. \ref{fig:SRFRvSIG},][]{ATLAS3DVII}. The data at high densities are dominated by galaxies in the core of the Virgo cluster (the densest LPE probed in \A3D). Unlike the original $T-\Sigma$ relation which was composed entirely of cluster galaxies from 55 rich clusters, the $kT-\Sigma$ relation of \citet{ATLAS3DVII} includes only one spiral-rich unrelaxed cluster, Virgo. We are thus almost completely ignorant of the kinematic-morphology density relation in clusters like those used to derive the original $T-\Sigma$ relation.

In light of the \fsr increase at the highest densities probed by \A3D, and the small number of clusters in the $kT-\Sigma$ relation, it is important to study the $kT-\Sigma$ relation of other clusters to investigate why the SR fraction suddenly increases in the core of Virgo and whether the SR fraction is truly independent of LPE density or GHE density (the average density of the host environment). \citet{DEugenio2013} performed an IFU survey of Abell~1689 at $z=0.183$ to investigate the $kT-\Sigma$ relation for one of the most massive and densest clusters known. Using the multiplexed IFU capability of the ESO FLAMES/GIRAFFE instrument, they identified that the average SR fraction of Abell~1689 is the same as for Virgo (no change with GHE) and that the fraction of SRs is enhanced in the densest regions and depleted in the lowest density regions of both clusters (i.e. a trend with LPE). Dynamical friction was proposed as an explanation for the segregation and the sudden increase in the SR fraction in the \A3D $kT-\Sigma$ diagram. However the degree of segregation in Abell~1689 is stronger than in Virgo. Both clusters are extremes: Abell~1689 is one of the most massive clusters known, and Virgo an unrelaxed low-mass cluster. Furthermore, current statistics on SRs are poor because they are rare. 

We present integral field spectroscopy (IFS) of \ngal galaxies in the Coma cluster \citep[z=0.024,][]{HanMould1992} observed with the Oxford SWIFT spectrograph \citep{SWIFT} to study the SR fraction and the SR segregation in a more typical cluster. We take special care to sample the galaxies without significant bias in luminosity or ellipticity which are known to affect the SR fraction directly. Throughout this work, we adopt a WMAP7 Cosmology \citep{WMAP7COSMO}; specifically, we use $H_{0}=70.4$\kms Mpc$^{-1}$, $\Omega_\n{m}=0.273$ and $\Omega_{\Lambda}=0.727$. All quoted uncertainties are the standard 68\% confidence interval (CI) unless otherwise stated. The structure of this paper is as follows: \S\ref{sec:dataanal} discusses the choice of the sample, the observations, the data reduction and the analysis techniques; \S\ref{sec:uncertainties} discussed uncertainties in derived quantites; \S\ref{sec:results} presents the kinematic maps, the \lameps diagram for the Coma cluster and an updated $kT-\Sigma$ relation for Virgo, Coma and Abell~1689; \S\ref{sec:discussion} discusses the context of the results and \S\ref{sec:conclusion} concludes. 

\section{Data and Analysis}
\label{sec:dataanal}

\subsection{Sample selection}
\label{sec:sample}

A complete survey of ETGs in a cluster is very observationally demanding, leading us to attempt inference from samples. Preliminary investigation revealed that \fsr could be estimated with an uncertainty of less than 10\% from a sample of 30 galaxies: approximating the distribution of SRs as binomial, we wish to infer \fsr (the probability of `success') in a sample of $n$ galaxies (the number of `trials'); if the true fraction is \fsr, then the intrinsic variance for the number of SRs (around a mean $nf_\textrm{SR}$) is simply $n f_\textrm{SR}(1-f_\textrm{SR})$ and the standard deviation about \fsr is $\sqrt{f_\textrm{SR}(1-f_\textrm{SR})/n}$. Assuming $f_\textrm{SR}=20$\% in a population of 30 galaxies, this approximation yields an inherent scatter of 7\%; such uncertainty is sufficient to distinguish between the core of Virgo (with \fsr$\sim30$\%) and the average population (with \fsr$\sim$15\%) to 95\% confidence. A more rigorous application of sample statistics is presented later in \S\ref{sec:errs:sample}, where we correctly model uncertainties (without replacement) using a hypergeometric distribution; the assumption of a binomial distribution here is conservative as it \emph{over} estimates the scatter in a sample (by a factor $\sqrt{(N-n)/(N-1)}$ where $N$ is the total number of galaxies in the population).

Without multiplexing, a survey of 30 galaxies is still demanding and requires observations spread over multiple semesters. The early results \citep[described in][]{Scott2012} consisted of IFS observations of 14 spectroscopically confirmed red sequence members of the Coma cluster. However, the selection was not representative of the parent luminosity or ellipticity distributions; rather they were chosen to have uniform representation in the logarithm of the galaxy velocity dispersion for greater diversity (wider sampling of the mass range). While this approach is useful for studying scaling relations \citep[e.g. the fundamental plane][]{DjorgovskiDavis87,Dressler87}, it was the limiting factor in our ability to determine \fsr in the Coma cluster: the sampling uncertainty far outweighed the random uncertainty in our measurements and the bias to higher luminosities was difficult to correct for. To determine \fsr more accurately we required a more representative sample, and as the sampling uncertainty drops to reasonable levels once the number of galaxies reaches around 30, we required another 16 galaxies, but chosen to alleviate bias. This `second sample', together with the sample of \citet{Scott2012}, is main subject of this paper. 

Starting with SDSS sources within a \sampradius\ radius of the cluster centre (20000 objects), we cross-matched all NED\footnote{NASA Extragalctic Database} galaxies with redshifts within \velrange of $z=0.024$ to create a spectroscopically confirmed sample of Coma cluster galaxies (377 objects). We then isolated the red sequence (see \S\ref{sec:data:photom}) in both this sample and the complete SDSS sample (over the same area of sky). Comparing the luminosity distributions of the red sequences in both samples suggested that the spectroscopic data in NED was 50\% complete at \nedspeccomplete\ mag. After forcing the selection of the \citeauthor{Scott2012} sample, we selected a sample from the spectroscopic catalogue to match both the luminosity function and the ellipticity distribution of the parent sample (individual galaxies were chosen randomly, subject to a few practical constraints such as ETG morphology and non-interaction with neighbours). 

\begin{figure}    \centering
   \includegraphics[width=0.5\textwidth]{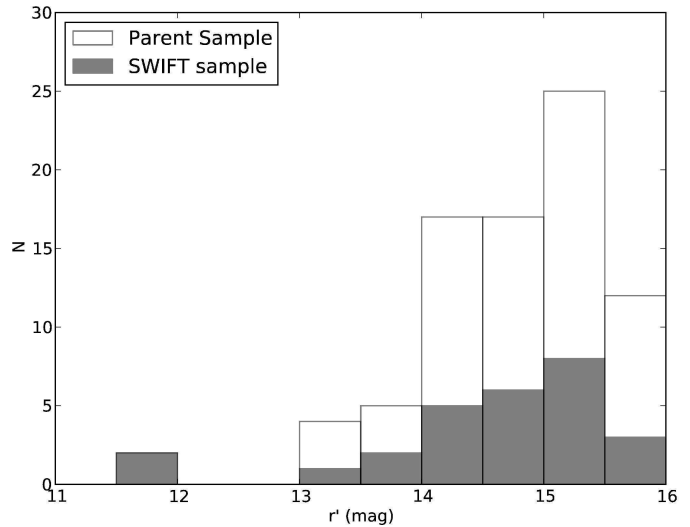} 
   \caption{The SDSS \rprime apparent magnitudes of the parent and sample populations; there is little or no bias evident in the luminosity function of the sample compared to the parent population. Given that we know that more luminous galaxies are more likely to be slow rotators, it is important for our sample to match the luminosity function of the cluster to avoid introducing bias into the derived \fsr}
   \label{fig:samplelumfunc}
\end{figure}

\begin{figure}   \centering
   \includegraphics[width=0.5\textwidth]{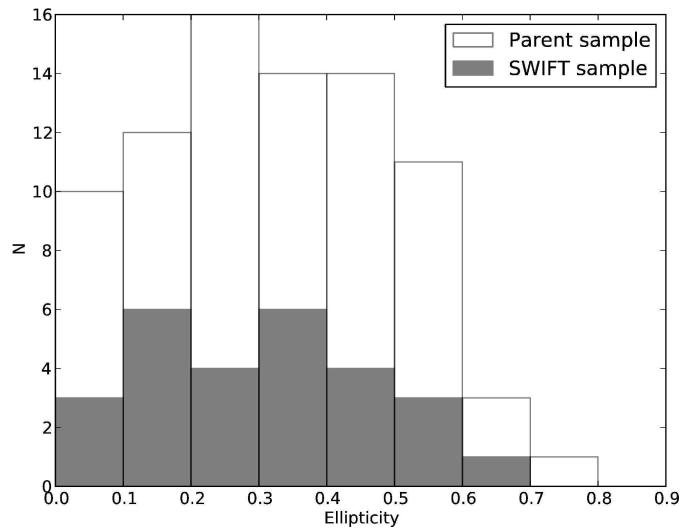} 
   \caption{The SDSS \rprime de Vaucouleurs model ellipticites for the parent and sample populations; there is little or no bias evident in the ellipticity distributions of the sample compared to the parent population. Given that we know of no fast rotators with \eps$>0.4$, it is crucial that our sample match the parent distribution of ellipticity to avoid introducing bias in the derived \fsr.}
   \label{fig:sampleellipdist}
\end{figure}

\begin{table*}
\caption{Details of individual Coma galaxies in the sample. Colours, magnitudes, ellipticities and effective radii were measured using SDSS {\sc Montage} images, except for \MK which was measured from the 2MASS {\sc Montage} image. The \grcol colour is measured inside a 3\farcs2 diameter; other magnitudes are Kron measurements. The I.D. is taken from \citet{Godwin1983}. The specific angular momentum \lam was measured from the SWIFT kinematic maps as described in the text; the value in parentheses gives the fraction of \Re over which the calculation was performed. The uncertainty in \lam is the formal (random) uncertainty (see \S\ref{sec:appendix:randomerror}). The rotator class, C is defined as 0 for FRs, 1 for galaxies with \lam below \SRdiv and 2 for morphologically identified SRs; a value of 3 corresponds to galaxies that are both classes 1 and 2. The probability of any galaxy being a SR is given by $p(\textrm{SR})$ and is described in \S\ref{sec:errs:obs}. The typical S/N in the binned SWIFT spectra (per 1\AA\ pixel) for each galaxy is given in the last row. }
\begin{center}
\small
\begin{tabular}{ccr@{$\pm$}lr@{$\pm$}lr@{$\pm$}lr@{$\pm$}lcr@{$\pm$}l@{~}lccc}
 I.D.  & $\Sigma_{3}$ &\multicolumn{2}{|c|}{ \gprime-\rprime} & \multicolumn{2}{|c|}{$M_{r^{\prime}}$ }& \multicolumn{2}{|c|}{$M_{K}$ }&  \multicolumn{2}{|c|}{\eps} & \Re & \multicolumn{3}{|c|}{\lam} & C & $p(SR)$ & S/N \\ 
 (GMP)  & kpc$^{-2}$ & \multicolumn{2}{|c|}{ mag} & \multicolumn{2}{|c|}{ mag} & \multicolumn{2}{|c|}{ mag} &  \multicolumn{2}{|c|}{ } & arcsec & \multicolumn{3}{|c|}{ } & & & \\
\hline
2390 &  84.3 &  0.92 & 0.01 & -21.46 & 0.002 & -24.46 & 0.006 & 0.214 & 0.002 &  14.3 & 0.16 &   0.03 & (0.2) &  0 &  0.22 & 41 \\ 
2457 & 113.1 &  0.87 & 0.02 & -19.49 & 0.004 & -22.34 & 0.016 & 0.416 & 0.026 &   3.1 & 0.38 &   0.02 & (1.0) &  0 &  0.00 & 18 \\ 
2551 & 139.0 &  0.92 & 0.02 & -20.11 & 0.003 & -22.92 & 0.013 & 0.452 & 0.004 &   7.7 & 0.35 &   0.01 & (1.0) &  0 &  0.00 & 14 \\ 
2654 & 180.2 &  0.93 & 0.01 & -19.64 & 0.004 & -22.53 & 0.015 & 0.142 & 0.011 &   2.0 & 0.21 &   0.01 & (1.0) &  0 &  0.00 & 18 \\ 
2805 & 420.5 &  0.90 & 0.01 & -19.44 & 0.005 & -22.33 & 0.017 & 0.221 & 0.011 &   2.7 & 0.31 &   0.01 & (1.0) &  0 &  0.00 & 18 \\ 
2815 & 298.2 &  0.85 & 0.01 & -19.90 & 0.004 & -22.72 & 0.014 & 0.510 & 0.005 &   3.5 & 0.40 &   0.04 & (1.0) &  0 &  0.00 & 22 \\ 
2839 & 244.1 &  0.90 & 0.01 & -20.06 & 0.004 & -23.09 & 0.012 & 0.080 & 0.025 &   2.1 & 0.29 &   0.01 & (1.0) &  0 &  0.00 & 23 \\ 
2912 &  34.9 &  0.91 & 0.01 & -19.95 & 0.004 & -22.93 & 0.013 & 0.289 & 0.011 &   3.4 & 0.29 &   0.01 & (1.0) &  0 &  0.00 & 19 \\ 
2921 & 939.1 &  0.91 & 0.01 & -23.15 & 0.001 & -26.27 & 0.003 & 0.359 & 0.002 &  38.0 & 0.04 &   0.01 & (0.1) &  3 &  1.00 & 48 \\ 
2940 & 447.8 &  0.88 & 0.01 & -19.82 & 0.004 & -22.70 & 0.014 & 0.066 & 0.014 &   2.7 & 0.32 &   0.03 & (1.0) &  0 &  0.00 & 10 \\ 
2956 &  80.6 &  0.99 & 0.01 & -20.00 & 0.004 & -22.96 & 0.012 & 0.640 & 0.004 &   3.9 & 0.50 &   0.01 & (1.0) &  0 &  0.00 & 18 \\ 
2975 & 943.0 &  0.86 & 0.01 & -21.41 & 0.002 & -24.32 & 0.007 & 0.022 & 0.003 &   7.4 & 0.09 &   0.01 & (1.0) &  2 &  0.00 & 19 \\ 
3073 &  87.0 &  0.91 & 0.01 & -20.63 & 0.003 & -23.65 & 0.009 & 0.151 & 0.007 &   4.7 & 0.22 &   0.02 & (1.0) &  0 &  0.00 & 17 \\ 
3084 &  34.8 &  0.92 & 0.01 & -19.61 & 0.004 & -22.48 & 0.016 & 0.124 & 0.009 &   2.8 & 0.13 &   0.01 & (1.0) &  0 &  0.07 & 17 \\ 
3178 &  83.2 &  0.92 & 0.01 & -19.87 & 0.004 & -22.74 & 0.014 & 0.299 & 0.010 &   2.9 & 0.24 &   0.01 & (1.0) &  0 &  0.00 & 12 \\ 
3254 & 759.9 &  0.86 & 0.02 & -19.20 & 0.005 & -22.20 & 0.019 & 0.300 & 0.017 &   2.8 & 0.32 &   0.02 & (1.0) &  0 &  0.00 & 13 \\ 
3329 & 756.5 &  0.95 & 0.01 & -22.93 & 0.001 & -25.93 & 0.003 & 0.115 & 0.002 &  50.4 & 0.08 &   0.02 & (0.2) &  3 &  0.92 & 33 \\ 
3352 & 574.4 &  1.05 & 0.01 & -20.67 & 0.003 & -23.72 & 0.009 & 0.082 & 0.011 &   5.0 & 0.30 &   0.02 & (1.0) &  0 &  0.00 & 25 \\ 
3367 & 356.4 &  1.01 & 0.01 & -20.76 & 0.003 & -23.69 & 0.009 & 0.230 & 0.006 &   5.9 & 0.41 &   0.03 & (1.0) &  0 &  0.00 & 24 \\ 
3423 & 116.0 &  0.99 & 0.01 & -20.32 & 0.004 & -23.35 & 0.011 & 0.429 & 0.020 &   2.5 & 0.41 &   0.01 & (1.0) &  0 &  0.00 & 23 \\ 
3433 & 147.7 &  0.90 & 0.02 & -19.35 & 0.005 & -22.27 & 0.017 & 0.318 & 0.020 &   2.3 & 0.16 &   0.02 & (1.0) &  1 &  0.68 & 13 \\ 
3522 & 257.9 &  0.94 & 0.01 & -19.54 & 0.005 & -22.34 & 0.017 & 0.127 & 0.023 &   1.7 & 0.26 &   0.01 & (1.0) &  0 &  0.00 & 19 \\ 
3639 & 185.2 &  0.94 & 0.01 & -20.80 & 0.003 & -23.82 & 0.009 & 0.285 & 0.012 &   3.1 & 0.24 &   0.02 & (1.0) &  0 &  0.00 & 26 \\ 
3792 &  74.5 &  0.99 & 0.01 & -21.44 & 0.002 & -24.54 & 0.006 & 0.162 & 0.003 &   7.5 & 0.07 &   0.04 & (1.0) &  3 &  0.89 & 34 \\ 
3851 & 105.8 &  0.92 & 0.02 & -19.00 & 0.005 & -21.73 & 0.022 & 0.226 & 0.011 &   2.3 & 0.31 &   0.02 & (1.0) &  0 &  0.00 & 12 \\ 
3914 &  69.0 &  0.93 & 0.01 & -19.50 & 0.005 & -22.33 & 0.017 & 0.121 & 0.043 &   1.3 & 0.26 &   0.02 & (1.0) &  0 &  0.00 & 15 \\ 
3972 &  66.2 &  0.91 & 0.01 & -19.61 & 0.004 & -22.39 & 0.016 & 0.187 & 0.026 &   2.2 & 0.18 &   0.01 & (1.0) &  0 &  0.00 & 13 \\ 
\tiny

\end{tabular}
\end{center}
\label{tab:completesample}
\end{table*}

\subsection{Photometric Observations}

We made use of the NASA/IPAC {\sc Montage} service\footnote{http://montage.ipac.caltech.edu} to mosaic SDSS images into a 1\degree$\times$1\degree\ image, centred on the NED coordinates for the Coma cluster. The 2MASS images, resampled to the SDSS plate scale of 0\farcs4, required a zeropoint correction of $5\log(0.4)$ mag to account for the change in scale. No further data reduction or cosmic ray removal steps were necessary.

\subsection{Spectroscopic Observations}

The Short Wavelength Integral Field specTrograph (SWIFT) mounted on the 200'' (5m) Hale telescope at the Palomar Observatory was used to observe a total of \ngal Coma galaxies (time constraints prevented us from observing 30 galaxies). Observations were made on the four separate observing runs on the 3rd--4th of May 2009, the 25th--26th March 2010, the 5th June 2010 and the 9th--14th May 2012. Table \ref{tab:completesample} tabulates properties for the individual galaxies in the sample. Fig. \ref{fig:samplelumfunc} illustrates the SDSS \rprime apparent magnitudes of the parent and sample populations; there is little or no bias evident. A \KS test reports the probability of both samples being drawn from the same distribution to be \pmagdist. Similarly, Fig. \ref{fig:sampleellipdist} illustrates the SDSS \rprime de Vaucouleurs model ellipticites for the parent and sample galaxies; again, there is little or no bias present and a \KS test reports the probability of both samples being drawn from the same distribution to be \pepsdist. 

\subsection{Data reduction}

The SWIFT data were reduced using SWIFT data reduction pipeline, written for {\sc IRAF} \citep[][in preparation]{SWIFTPIPE}. 

The pipeline includes standard CCD data reduction steps such as bias subtraction, wavelength calibration and flat fielding as well as IFU specific stages such as cube reconstruction and illumination correction. The final wavelength calibration is accurate to better than ~0.1\AA. Cosmic rays are removed using the \LACOSMIC\ routine \citep{LACOSMIC}. We correct for flexure along the spectral axis using the night sky emission lines. Sky emission was removed to first order (see \S\ref{sec:data:kin}) by subtracting sky frames adjacent in time to each science exposure; targets were either observed using the standard NIR `ABBA' technique or using dedicated sky frames when the target occupied the full instrument \fov. Data cubes were combined using a dedicated python code, using offsets derived from the galaxy centroid in the wavelength-collapsed cubes. Although telluric standards were observed along with the science observations, no significant telluric absorption is present at the wavelengths used to calculate the kinematics so we do not attempt to correct for it. 

\subsection{Data analysis}

\subsubsection{Photometry}
\label{sec:data:photom}

Integrated photometry was measured directly from the SDSS and 2MASS {\sc Montage} images in \gprime, \rprime and $K$ using \sextractor (v2.5.0). When calculating the \grcol, we used apertures of diameter 3\farcs2. Kron magnitudes were adopted as total magnitudes (for use in the colour magnitude diagram and calculation of \sig3). The same images were also used for variance/weight maps in \sextractor. Detailed masks were created to obscure bright stars and their diffraction spikes.

\begin{figure}
   \centering
   \includegraphics[width=0.5\textwidth]{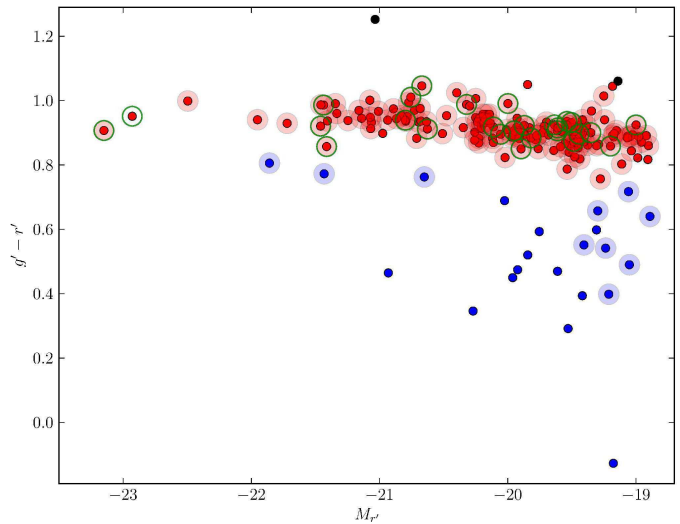} 
   \caption{The CMD of the Coma cluster. Photometry was measured from the {\sc Montage SDSS} images using {\sc Sextractor}. The RS was isolated using the techniques described in \citet{Houghton2012}; red points refer to galaxies in the 95\% CI of the derived CMR, while black and blue points signify `outliers' above and below the RS, respectively. Points with halos are spectroscopically confirmed in NED. Green circles highlight the \ngal SWIFT sample.}
   \label{fig:CMD}
\end{figure}

Using the \sextractor catalogues, we fit a double Gaussian mixture model to the red sequence (RS) and outlier distribution \citep[using MCMC techniques as described in][]{Houghton2012}; this allows us to isolate red sequence galaxies in the cluster \citep[the techniques used to clean the catalogues of stars and bad photometry are also described in][]{Houghton2012}. We chose an apparent magnitude limit of $r^{\prime}<$16.2 mag when fitting the colour magnitude relation (CMR); a limit around 16 mag was desirable because this study aims to be comparable to \A3D (with \MK$<-21.5$~mag) but we found that \sextractor Kron magnitudes are slightly fainter ($\sim0.2$~mag) than the SDSS model magnitudes used in the selection process (\S\ref{sec:sample}); such systematic differences are to be expected \citep{GrahamDriver2005}.  All objects within the 95\% CI of the derived CMR parameters were defined as \emph{red} galaxies on the RS; `outliers' above and below the RS were defined as `extremely red' or `blue', respectively. The resulting colour magnitude diagram (CMD) is shown in Fig. \ref{fig:CMD}. No spectroscopic information was used to correct for contamination by interlopers, although we do highlight spectroscopically confirmed cluster members (quoted by NED, see \S\ref{sec:sample}) in Fig. \ref{fig:CMD}.

We derive the local surface density using all reliable detections (extremely red, red and blue) with \rprime$<$16.2 mag. We define \sig3 to be three times the reciprocal of the smallest circular area ($A_{3}$, measured in kpc$^{2}$ at the distance of Coma) that encloses the nucleus of the 3rd nearest neighbour (with $M_\textrm{K}<-21.5$mag). We must also make a correction for foreground/background galaxies which are not at the redshift of Coma. Using surveys of galaxy number counts \citep{Yasuda2001}, we estimate $2.4\times10^{-3}$ galaxies per square arc min for $r^{\prime}<16$~mag. Thus 
\begin{equation}
\Sigma_{3} = \frac{3 - (1.7\times10^{-4}) a_{3}}{A_{3}}
\end{equation}
where $a_{3}$ is the same as $A_{3}$ but measured in square arc minutes. Note that we do not apply a constraint on the line-of-sight velocity as we have insufficient information for all photometric objects and furthermore, the velocity dispersion of the cluster is so high ($\sim1000$ \kms) that the usual constraint ($\Delta V_\textrm{los}<300$\kms) would be unsuitable. However, when considering the ETG population, as we do in \S\ref{sec:results}, we take as a proxy all galaxies on the RS. We have not visually confirmed all red objects with \rprime$<16.2$ mag to be ETGs. However the SWIFT ETG sample was morphologically verified by eye, using the SDSS images. It is reassuring to see that all the SWIFT sample lies entirely on the RS in Fig. \ref{fig:CMD}, adding confidence to the assumption that the RS traces ETGs. Clearly, this assumption may not be true in a field sample, but for a well evolved, low redshift cluster like Coma, there are very few exceptions, particularly in the central 15\arcmin as relevant to this study.  

Surface photometry (\eps, \Re and \mue) was measured directly from the \rprime SDSS {\sc Montage} image. We integrated the pixel counts in circular apertures outwards from the centre and fitted the integral of the de Vaucouleurs profile to this curve of growth \citep[described in][]{Houghton2012} to determine the effective radius \Re and average effective surface brightness \mue. The ellipticity \eps was measured in a similar way to the SAURON and \A3D surveys: within the elliptical isophote of area $\pi R_\textrm{e}^{2}$, we calculated the second moments and the corresponding ellipticity.  

\subsubsection{Stellar kinematics and binning}
\label{sec:data:kin}

We use the \ppxf software \citep[][v4.65]{PPXF} to calculate the stellar kinematics from the SWIFT spectra using the calcium triplet. We model the intrinsic line-of-sight velocity distribution of the galaxies with a Gaussian, parametrized by a first moment (velocity, $V$) and second moment (dispersion, $\sigma$). We provided \ppxf with the \citet{Cenarro2001} library of stellar spectra (covering the range $0.8348<\lambda(\mu m)<0.902$). Accurate initial guesses for the systemic velocity were crucial to extract reliable kinematics; these were estimated by eye for each galaxy using the calcium triplet. We fit kinematics typically in the range $0.86 < \lambda (\mu \textrm{m}) < 0.89$ with small variations around $0.05 \mu$m.

To optimise the extraction of kinematics across the galaxy, it was necessary to bin up spaxels to increase the signal to noise ratio (S/N). We chose a method based on azimuthal sectors \citep[e.g.][]{Nowak2007,Rusli2011}. After splitting the azimuthal angle into a predetermined number of divisions, we adaptively binned spaxels, working outwards in radius, to achieve the desired S/N limit. In order to maximise spatial coverage, is was desirable to bin spectra up to the largest isophote which fitted inside the (coadded) \fov. However, in practice this was not always possible due to significant sky emission residuals progressively dominating over the source flux at larger radii. Therefore the outer isophote level and corresponding radius were not fixed, but chosen individually for each galaxy: observations made under light cloud cover often exhibited larger sky residuals (both from strong and weak skylines), preventing us from binning out to larger and fainter isophotes. Similarly, the target S/N limit was not fixed: galaxies with a higher velocity dispersion required a higher S/N limit because the absorption features are shallower and more difficult to measure. Typical S/N limits ranged between 10 and 40 (per 1\AA\ pixel) and are listed in Table \ref{tab:completesample}. Spaxels within a radius of 0\farcs47 were binned to a single central aperture. 

The spaxels from SWIFT do not have identical spectral resolutions. In order to match the resolution of the stellar library with the resolution of the galaxy observations, we measured the spectral resolution of the binned spectra using the skylines; we independently fit Gaussians to seven skylines surrounding the observed wavelength of the calcium triplet and chose the median FWHM as the formal resolution for that bin. In all binned spectra, the skylines were well represented by Gaussian profiles; no asymmetries, wings or top-hat profiles were apparent. We used this resolution as the instrumental resolution when deriving the kinematics of each bin with \ppxf; typically $\sigma_\textrm{inst}$ varied between 45 \kms and 55 \kms across the field of view. Similarly, we also found deviations in the wavelength calibration of order a few \kms.

Exceptionally large sky line residuals can cause \ppxf to find false solutions (particularly with respect to velocity). Although we made use of the {\sc CLEAN} keyword in the \ppxf software to reject highly deviant pixels (with just a single iteration), this was insufficient to ensure a robust solution. For this reason, we investigated masking the sky lines or simultaneously fitting the sky spectrum with the kinematics \citep[as in][]{Weijmans2009}. This investigation is summarised in Appendix \ref{sec:appendix:skyerror}. We found that simultaneously fitting the sky emission with the kinematics gave the most robust results with no obvious failures; masking sky lines did almost as well, but failed on a few cases. 

While investigating the systematic uncertainty associated with the discretisation of the kinematic maps (see \S\ref{sec:lamerrs} \& \ref{sec:appendix:binningerror}), we changed the binning geometry (by rotating the radial divisions of the sector patten) and re-extracted the kinematics. Averaging these multiple realisations provides a smoother representation of the data (hereafter referred to as the \emph{dithered maps}), without formally smoothing the maps. Clearly the different kinematic realisations are correlated, but by perturbing the bin positions, we recover information on scales smaller than the size of the bins. This is best shown in the velocity map of GMP3423: in an individual realisation, the wide azimuthal angle of the bins disperses (azimuthally) the velocity map extremities (the maximal rotation along the major axis); in the dithered map, the maximal rotation curve is confined to a narrower azimuthal width along the major axis. The use of dithering to recover information on scales smaller than the sampling is well documented \citep[e.g. the {\sc drizzle} concept used to recover diffraction limited imaging from undersampled images on the HST,][]{Drizzle}. A full exploration of this technique is beyond the scope of this work, and merits a separate investigation on its own. However, we find use for the dithered maps in calculating \lam (see below). 

\subsubsection{Calculation of specific angular momentum, $\lambda$}

We calculate \lam following the approaches of the SAURON and \A3D surveys. Briefly summarised, we use a circular de Vaucouleurs curve of growth method to measure \Re from the SDSS photometry (\S\ref{sec:data:photom}), find the elliptical isophote with area equal to $\pi R_{e}^{2}$ and calculate the moment ellipticity $\epsilon_{e}$ and position angle $\phi_{e}$ of the light falling within that isophote. Within an ellipse defined by these \emph{effective} parameters $\epsilon_{e}$ and $\phi_{e}$ (again with area $\pi R_{e}^{2}$), centred on the first moment within the same aperture, we calculate \lam as per the normal expression \citep{ATLAS3DIII},
\begin{equation}
\label{eq:lam}
\lambda = \frac{\sum_{i}^{N}R_{i} F_{i} |V_{i}|}{\sum_{i}^{N}F_{i} R_{i} \left(V_{i}^{2}+\sigma_{i}^{2}\right)^{1/2}}
\end{equation}
where $R_{i}$, $F_{i}$, $V_{i}$ and $\sigma_{i}$ are the radius, flux, velocity and velocity dispersion of the $i$th element and $N$ is the number of elements enclosed within the $\{\epsilon_{e},\phi_{e},\pi R_{e}^{2}\}$ ellipse. We discuss the associated uncertainties in \S\ref{sec:uncertainties}.

Where the ellipse defined by the effective parameters encompasses only a fraction of a bin, $R_{i}$ and $F_{i}$ are the mean radius and total flux of the enclosed portion of that bin. When the ellipse defined by the effective parameters was larger than the extent of the kinematic information, we calculate \lam using the full map (we quote the fraction of $R_{e}$ covered by our maps in Table \ref{tab:completesample}). For such galaxies, we measure \lam and \eps on different scales. Comparison of data to the Virial Theorem (and its relatives) requires measurements to be made on the same scale (or mass fraction). Although this condition is not always met, for the few galaxies concerned \eps does not vary strongly with radius and so would not change significantly if it were measured over the aperture defined by the kinematics maps. 

We wish to classify ETGs as either fast or slow rotators. Most recently, this has been done with kinemetry \citep{Kinemetry,ATLAS3DII}. We investigate two approaches here. We can morphologically classify our data visually, like the SAURON survey, depending on whether the velocity map exhibits global, large scale rotation or not. This is similar to the kinemetry approach, but lacks well defined parametric limits. We also use the division in the \lameps plane: slow rotators appear to inhabit the region defined by $\lambda<0.31\sqrt\epsilon$ \citep{ATLAS3DIII}. However, the uncertainties in \lam and \eps may not be negligible, as discussed below, so careful modelling of random and systematic uncertainties is required to robustly make use of this approach, particularly at higher redshifts \citep[as was done in][]{DEugenio2013}.

\section{Estimates of uncertainty}
\label{sec:uncertainties}

We propagate uncertainties by approximating the Poisson photon statistics with a Normal distribution together with the first order (derivative) approach. This neglects covariances introduced by interpolation of the data and provides uncertainties in integrated magnitudes, aperture colours and kinematics\footnote{pPXF derives parameter uncertainties, based on the input uncertainties of the galaxy spectrum, using the Levenberg-Mardquart algorithm but it ignores the covariances with template mismatch and sky subtraction which are optimised outside of this algorithm.}. 

\subsection{Uncertainties in specific angular momentum \lam}
\label{sec:lamerrs}

In Appendix \ref{sec:appendix:lamerrs} we investigate different contributions to the uncertainty in \lam: the formal random uncertainty (from photon statistics) in deriving \lam from a single realisation (\S\ref{sec:appendix:randomerror}), the systematic uncertainty from discretisation of the kinematic maps (\S\ref{sec:appendix:binningerror}), and the systematic uncertainty in the kinematics originating from sky line residuals (\S\ref{sec:appendix:skyerror}). 

We find that the formal random uncertainty is typically the dominant source of uncertainty (we derive expressions for the first-order propagation of uncertainties in \S\ref{sec:appendix:randomerror}, which are not trivial). The discretisation error is nearly always smaller than the formal random uncertainty and can be minimised further by using the dithered maps to calculate \lam. However, the uncertainty from photon noise cannot be reduced by dithering so we adopt the \emph{average} random uncertainty in \lam calculated from single binning realisations (quoted in Table \ref{tab:completesample}). Furthermore, when we fit the sky spectrum simultaneously with the kinematics, the systematic uncertainty from the sky residuals is greatly reduced (typically $<0.01$ in \lam) which is smaller than the formal random uncertainty.

\subsection{Uncertainties in \eps}

When estimating \eps, we quote the RMS deviations from a polynomial fit to the ellipticity profile between $0.1<R/R_{e}<10$. 

\subsection{Uncertainties in \fsr}
\label{sec:errs:fsr}
There are two principle sources of uncertainty when we infer \fsr. Firstly, the accuracy of our observations leads to uncertainty in the number of slow rotators found in the sample; we call this measurement uncertainty. Secondly, there is uncertainty from using a finite sample; we call this sample uncertainty. We now discuss these uncertainties in the next two sections, followed by a discussion on how to combine them.

\subsubsection{Measurement uncertainties}
\label{sec:errs:obs}

We have calculated uncertainties for both \lam and \eps; we now wish to propagate these uncertainties into our calculation of \fsr. When \fsr is estimated using morphological analysis of the kinematic maps, such propagation is unclear. However, if one classifies galaxies with $\lambda < 0.31 \sqrt{\epsilon}$ as being slow rotators, the propagation of uncertainties can be approximated with Monte Carlo techniques, as in \citet{DEugenio2013}. 

Let us make the assumption that the true value of \lam and \eps for each galaxy is normally distributed around our measurements (with no correlation between \lam and \eps), using standard deviations defined by our uncertainty estimates. We must also truncate and renormalise the normal distributions to ensure that $0<\lambda<1$ and $0<\epsilon<1$. By sampling these distributions many times, we can infer the probability of any one galaxy having $\lambda < 0.31 \sqrt{\epsilon}$ (quoted in Table \ref{tab:completesample} as $p(\textrm{SR})$). Similarly, by sampling the distributions of all galaxies simultaneously, we can infer the probability of any number of the galaxies having $\lambda < 0.31 \sqrt{\epsilon}$; this is important if many galaxies individually have quite low probabilities of being slow rotators, because the probability of any one of them being a SR may be significant.

\subsubsection{Sample uncertainties}
\label{sec:errs:sample}

To put results from our sample into context, we should define the uncertainties associated with inferring quantities from a (representative) subset of the population and similarly the uncertainties when inferring from the \emph{complete} population. In the latter case, it is not true that complete samples are free from uncertainty: a finite sample drawn from a binomial distribution will always show variation in the number of `successes' because the variance of that binomial distribution is non-zero. The type of uncertainty we choose to quote depends on the question we wish to answer; in the case of a subset of galaxies taken from Coma the obvious question is ``How many slow rotators are there likely to be in Coma (given a subset of galaxies with certain constraints on luminosity and environment)?''. Thus here, we wish to `correct' an analysis to infer the intrinsic value of some parameter \emph{in the parent sample}. Conversely, we may ask the question ``Is the intrinsic \fsr of this population the same as other populations (given constraints on luminosity and environment)?''. This is subtly different to the former question, but the solutions are similar as we now show. 

Given a sample of $n$ galaxies taken from population of $N$ galaxies, let us attempt to infer the \emph{true} number of slowly rotating galaxies in the population $K$, given that we observed in our sample only $k$ slowly rotating galaxies. Formally, the hypergeometric distribution tells us the probability of finding $k$ SRs in a sample of $n$ from a parent population of $N$ galaxies which actually has $K$ slow rotators
\begin{equation}
\label{eq:hypergeo}
p(k | n,K,N) = \frac{\left(^{K}_{k}\right) (^{N-K}_{n-k})}{(^{N}_{n})}
\end{equation}
where $(^{a}_{b})$ is the binomial coefficient. To estimate the uncertainty on $K$ we must use Bayes Theorem,
\begin{equation}
\label{eq:bayes}
p(K | k,n,N) = \frac{ p(K | n,N) p(k | n,K,N)} {p(k | K,N)}.
\end{equation}
The term $p(K | n,N)$ is the prior. We use uninformative flat priors for $K$, allowing with equal probability $0\leq K \leq N$. Furthermore, we calculate the denominator in Eq.~\ref{eq:bayes} by requiring that $p(K | k,n,N)$ is normalised. 

We can now estimate the uncertainty in the true number of SRs in the Coma cluster given our selection criteria, and with (or without) additional luminosity or environment constraints. In any given bin of luminosity or projected density, we only need to know the sample size $n^\prime$, the observed number of slow rotators in that bin $k^\prime$ and the actual number of galaxies from the cluster that fall into that bin $N^\prime$ to estimate the uncertainty in the true number of SRs $K^\prime$ using Eq. \ref{eq:bayes}. However, in this way we estimate the uncertainty (actually the posterior) on the number of SRs only in \emph{Coma}, given our selection criteria and any other assumptions we have made. We are \emph{not} estimating the posterior for the number of SRs in the galaxy population as a whole, or even for clusters `like Coma'. If we wished to do this, we should replace the hypergeometric distribution with the binomial, with a probability of `success' $p=K/N$. Naturally, this is the large $N$ limit when using the hypergeometric distribution for finite populations. Therefore, even for complete surveys such as \A3D, there is a posterior uncertainty associated with inferring \fsr for \emph{all} ETGs in the Universe at $z\sim0$, or for inferring \fsr in clusters \emph{like} Virgo. We plot this uncertainty on the \A3D data in Figs. \ref{fig:SRFRvMK} \& \ref{fig:SRFRvSIG}. 

\subsection{Combining measurement and sample uncertainties}
\label{sec:errs:comb}
We know that our measurements are uncertain and that inference from a sample (whether sparse or complete) also carries an uncertainty. If one were significantly larger than the other, we could neglect the smaller, but unfortunately that is not the case here. To reliably report on \fsr in the Coma cluster requires us to account for both these sources of uncertainty. 

The technique described in \S\ref{sec:errs:sample} provides $p(K | k, n, N)$, which seems to provide an answer to how many SRs there are in the parent population. However, if our assumption about the value of $k$ was incorrect, then this answer is also incorrect. The discussion in \S\ref{sec:errs:obs} highlights that our measurement of $k$ is uncertain, and it describes how to estimate a \emph{probability distribution} $p(k | n, N)$ for the number of SRs found in our sample. With this information, we can marginalise over all possible values of $k$ to recover $p(K | n, N)$. Formally,
\begin{equation}
p(K | n, N) = \sum\limits_{i} p(K | k_{i}, n, N) p(k_{i} | n, N)
\end{equation}
where $i$ indexes the possible values of $k$. If there were significant uncertainty in $n$ and $N$, we could similarly integrate them out.

For Abell~1689, we can also combine the measurement uncertainty with the sample uncertainty, as $p(SR)$ is known for each galaxy. For the \A3D data however, uncertainties on \lam and \eps were not published, so we can only quote $p(K | k, n, N)$ for these data in in Figs. \ref{fig:lam-eps} \& \ref{fig:SRFRvMK}; such uncertainties are likely to be underestimates.

\section{Results}
\label{sec:results}

\begin{figure*}
   \centering
   \includegraphics[width=0.9\textwidth]{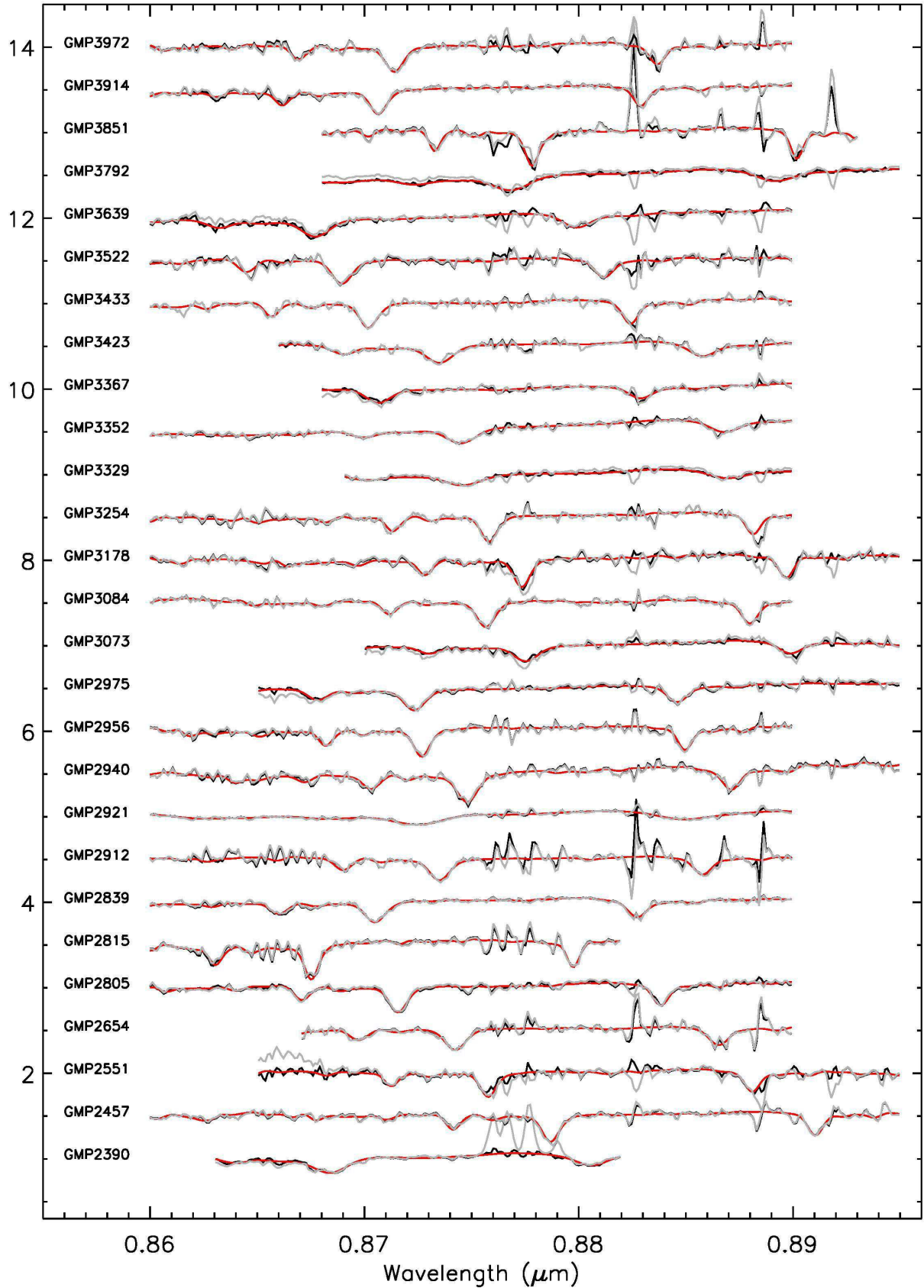} 
   \caption{Typical spectra for the \ngal galaxies in this sample. The spectra have not been corrected for recession velocity and are sorted by GMP number. Galaxy spectra in black are shown with optimised sky subtraction while galaxy spectra in grey show typical residuals using an first order (standard) sky subtraction. The best fitting kinematic model spectrum is shown in red for each spectrum. Spectra have been normalised by the (median) continuum level and are spaced vertically by intervals of 0.5. }
   \label{fig:megaspec}
\end{figure*}

Typical (binned) SWIFT spectra for each Coma galaxy are shown in Fig. \ref{fig:megaspec}. These examples illustrate the median quality (in terms of S/N) for each galaxy. They have not been corrected for individual recession velocities, allowing the quality of the sky subtraction (the dominant source of systematic error, see \S\ref{sec:appendix:lamerrs}) to be compared at each wavelength. We overplot the best fitting kinematic model spectrum in red. 

We present kinematic maps for the sample of 27 Coma cluster ETGs in Fig. \ref{fig:kina}. For each galaxy, the reconstructed image is shown with the dithered velocity and velocity dispersion maps. The colour bar scales for both the velocity and dispersion maps are fixed; this is to highlight the differences between the galaxies.

\begin{figure}
   \centering
   \includegraphics[width=0.5\textwidth]{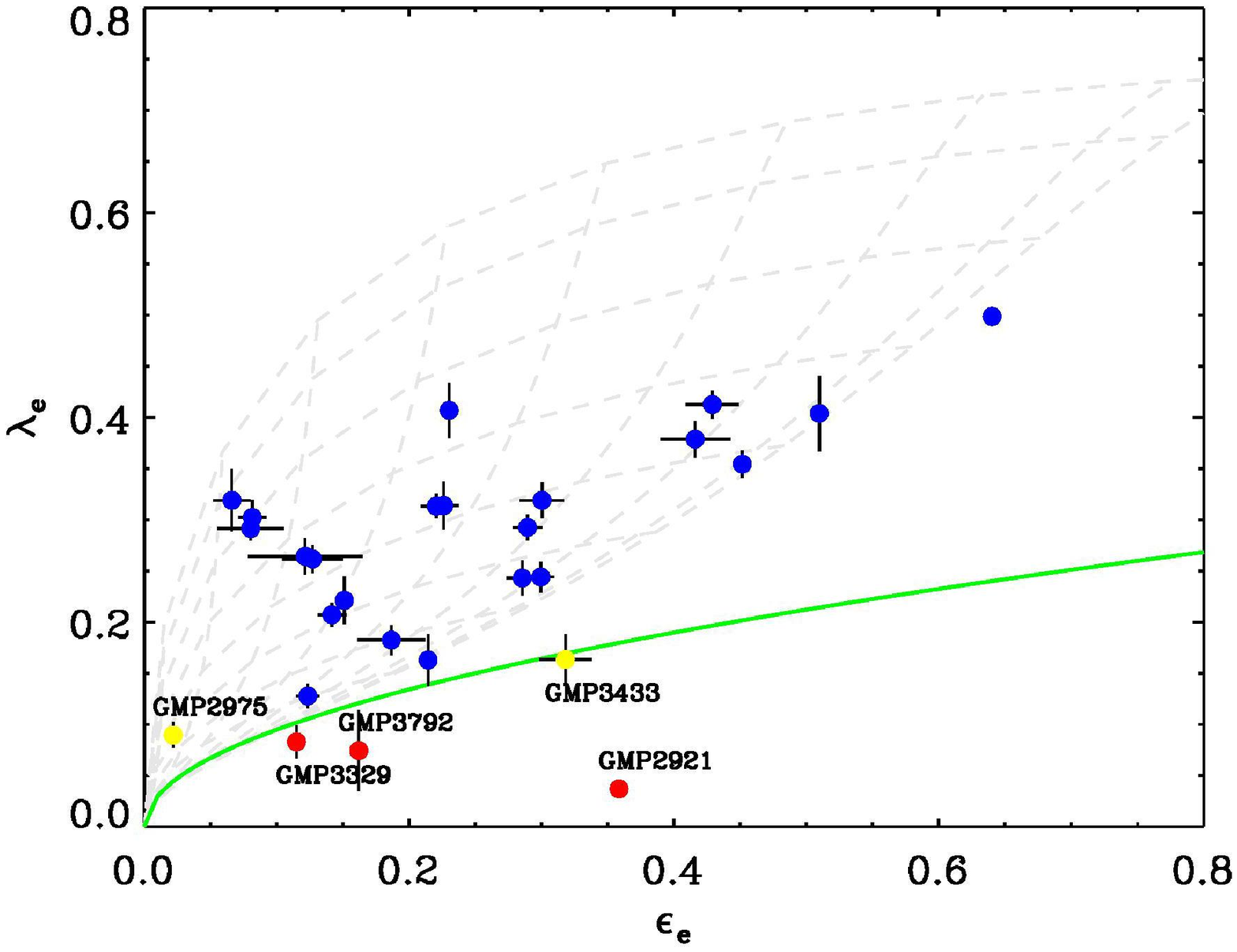} 
   \caption{The $\lambda$--$\epsilon$ plot for the Coma cluster. Errors show random uncertainty only. The light grey grid projects oblate ellipsoid models (of increasing flattening and anisotropy) at various inclinations. The green line (\SRdiv) represents a formal division between (projected) fast rotators and slow rotators in the \A3D survey \citep{ATLAS3DIII}. We label all types of SRs; class 1 or 2 are coloured yellow while class 3 are coloured red. FRs are coloured blue.}
   \label{fig:lam-eps}
\end{figure}

Fig. \ref{fig:lam-eps} shows the \lameps diagram for our Coma sample. If we select SRs morphologically based on the absence of significant rotation we find four examples: GMP2921, GMP3329, GMP3792 and GMP2975. If we classify SRs as all galaxies with \lam$<$\SRdiv, then there are also four examples in Fig. \ref{fig:lam-eps}: GMP2921, GMP3329, GMP3792 and GMP3433. Using the Monte Carlo techniques described in \S\ref{sec:errs:fsr}, we formally find \NSR with \lam$<$\SRdiv. 

There are three galaxies that satisfy both the morphological and the \lameps constraints (red in Fig. \ref{fig:lam-eps}) and two galaxies that satisfy one or the other (yellow in Fig. \ref{fig:lam-eps}). Table \ref{tab:completesample} summarises these results. Thus however we classify a SR, we find \fsr in the ETG population of Coma to be around 15\%. \citet{ATLAS3DIII} highlight that the \lameps division is only representative and wherever possible, it is preferable to classify galaxies individually based on the appearance of the velocity map. Accordingly, hereafter we report on the number of morphological SRs ($\textrm{C}\geq2$ in Table \ref{tab:completesample}) but we derive corresponding uncertainties from the \lameps diagram using the techniques in \S\ref{sec:errs:fsr}.

GMP2921 and GMP3329 show no evidence of rotation and are undisputed SRs. Similarly GMP3792 and GMP2975 show no evidence of global rotation extending to large radii. All four of these morphological SRs have $\lambda<0.1$. Although GMP2975 is consistent with being a fast rotator in the \lameps diagram (\eps=0.022, \lam=0.08), suggesting it is an oblate nearly isotropic ellipsoid seen face on, it does not show global rotation in its velocity map. Were it a FR, the necessary alignment along the line of sight would be extremely rare; in \A3D (260 galaxies), no fast rotator with $\{\lambda,\epsilon\}<0.1$ was found \citep{ATLAS3DIII}. Assuming GMP2975 is a SR, either (underestimated) measurement error scattered it above \SRdiv, or slow rotators are occasionally found above this fast/slow division.

The SRs we identify all have $\epsilon<0.4$, consistent with the SRs in \A3D. Conversely, all the FRs detected have $\lambda < 0.6$; a similar result was found in \citet{DEugenio2013}. However in \A3D the FRs populate up to $\lambda<0.8$. 

The majority of the dispersion maps in Fig. \ref{fig:kina} are blue (dynamically cold, $\sigma<100$\kms), while one dispersion map stands out from all the others as being dynamically much hotter than the others; that of GMP2921. This galaxy is likely the most massive in the cluster and is one of the two central dominant galaxies. However, the other centrally dominant galaxy, GMP3329, has a very moderate dispersion map in comparison; indeed the dispersion map of the non-central galaxy GMP3792 is just as dynamically hot. 

The dispersion maps do not universally show dynamically hotter components at their centres; indeed, they seem to show a range of radial gradients. GMP2457, GMP2551, GMP2975 and GMP3433 show no clear rise in dispersion towards their centres; conversely, GMP2839, GMP2912, GMP3423 and GMP3914 show clear evidence of a rise in dispersion. While the dispersion maps are in general noisier than the velocity maps, the strength of a central, dynamically hotter component can be judged in nearly all cases; the exceptions are GMP2390, GMP3073, GMP3352 and GMP3357 where noise in the maps is sufficient to prevent clear identification of radial gradients. 

We also find trends with the dispersion maps. Dynamically cold dispersion maps are associated with rotating velocity maps (FRs). Conversely, SRs are generally associated with dynamically hotter dispersion maps. However, dynamically hot dispersion maps are \emph{not} always SRs. A concern for this rule is GMP2975, which we classify morphologically as a SR despite its position in the (FR region of the) \lameps plot of Fig. \ref{fig:lam-eps}; although having a relatively cool dispersion map, it is not technically `cold' because $\sigma>100$\kms. Interestingly, there is no evidence of increasing dispersion towards the centre of this galaxy.

We present \fsr as a function of density ($\log\Sigma_{3}$) and luminosity ($M_\textrm{K}$) in Figs. \ref{fig:SRFRvSIG} \& \ref{fig:SRFRvMK}, respectively. As discussed in \S\ref{sec:uncertainties}, it is more useful to consider uncertainties on the posterior than on the likelihood, so we show the 68.2\% CI of the posterior probability. Posterior uncertainties are not the same as maximum-likelihood uncertainties. For example, in Fig. \ref{fig:SRFRvSIG}, for the lowest densities that we sample, there are only two galaxies in the bin and none are found to be slow rotators. In this case, the formal uncertainty on the posterior (we quote $\pm32.1\%$ from the median) does not encompass the actual observation: it is far more likely, given the very low number statistics and the uninformative priors, that the true number of SRs is above zero at this density. We could be informative here: we know \fsr in Abell~1689 is around 15\% \citep{DEugenio2013}; similarly, we know that SRs must have $\epsilon<0.4$, providing an upper limit to \fsr. But we do not investigate these informative priors. 

Fig. \ref{fig:SRFRvMK} confirms that SRs are more common than FRs at higher luminosities in all GHEs (field/group, Virgo, Coma or Abell~1689). We do not find any low luminosity slow rotators in our sample (all SRs have \MK$<24$~mag), although the finite probability of misclassifying these galaxies in the \lameps plot (Fig. \ref{fig:lam-eps}) leads the posterior to favour the presence of low luminosity SRs at the 68\% confidence level. Specifically, the low luminosity galaxy GMP3433 biases the posterior in this way; it is likely to originate from below \SRdiv, although we classify it morphologically as a FR due to the clear rotation gradient in its velocity map. The posterior would not have been in favour of any low luminosity SRs had we quoted the uncertainty from just a sample analysis $p(K | k, n, N)$.

Fig. \ref{fig:SRFRvSIG} shows that the \emph{average} \fsr in different GHEs is remarkably constant at around 15\%. A simple linear fit with uncertainties in \fsr alone\footnote{as described in \S15.2 of \citet{NRC}} provides a best fit of \fsr $=  (0.009 \pm 0.02) \log \Sigma_{3}  + (0.14 \pm 0.02)$ and suggests that the slope is consistent with zero. Within individual clusters, there is strong evidence for an increase in \fsr in denser LPEs and similarly, a decrease in less dense LPEs; the evidence is particularly strong for Abell~1689 \citep{DEugenio2013}. Note however that the low density LPE limit where only FRs are found is different for each cluster; while no SRs are found in Abell~1689 at a LPE density of $\log\Sigma_{3}\sim2.2$, the same LPE density in Virgo holds the highest \fsr in the entire cluster.  

\newpage

\begin{figure*}
   \centering
   \subfloat{\includegraphics[width=0.5\textwidth]{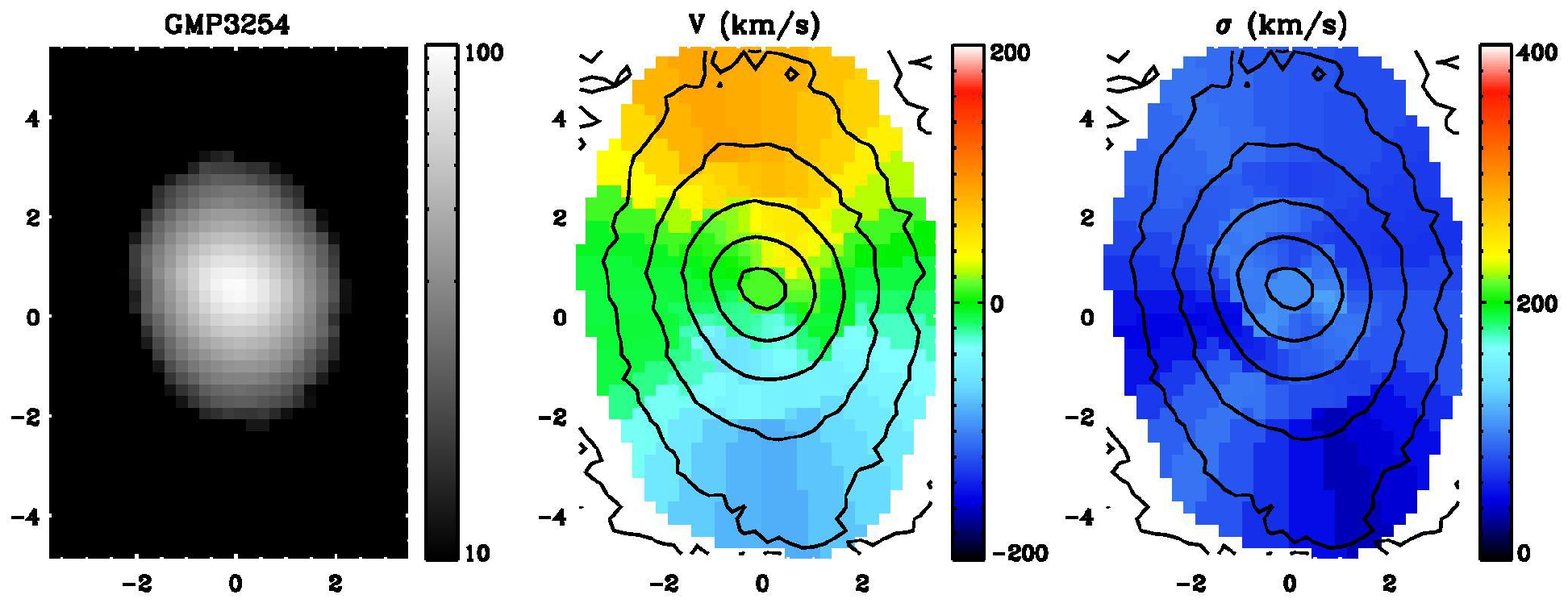}}      \subfloat{\includegraphics[width=0.5\textwidth]{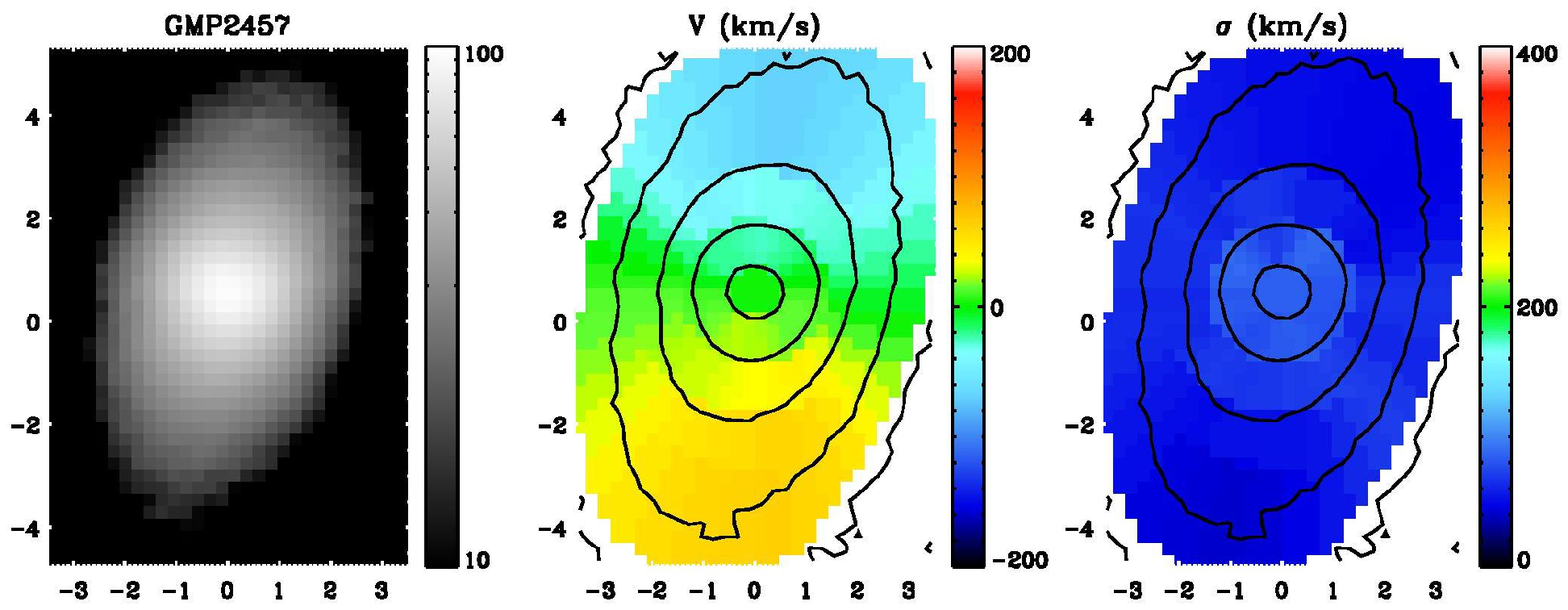}}
   
   \subfloat{\includegraphics[width=0.5\textwidth]{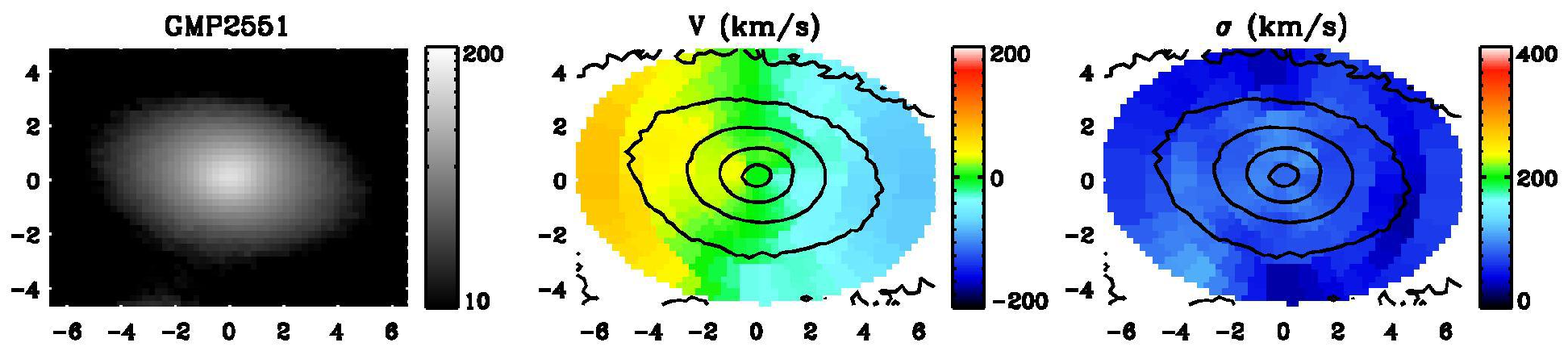}}      \subfloat{\includegraphics[width=0.5\textwidth]{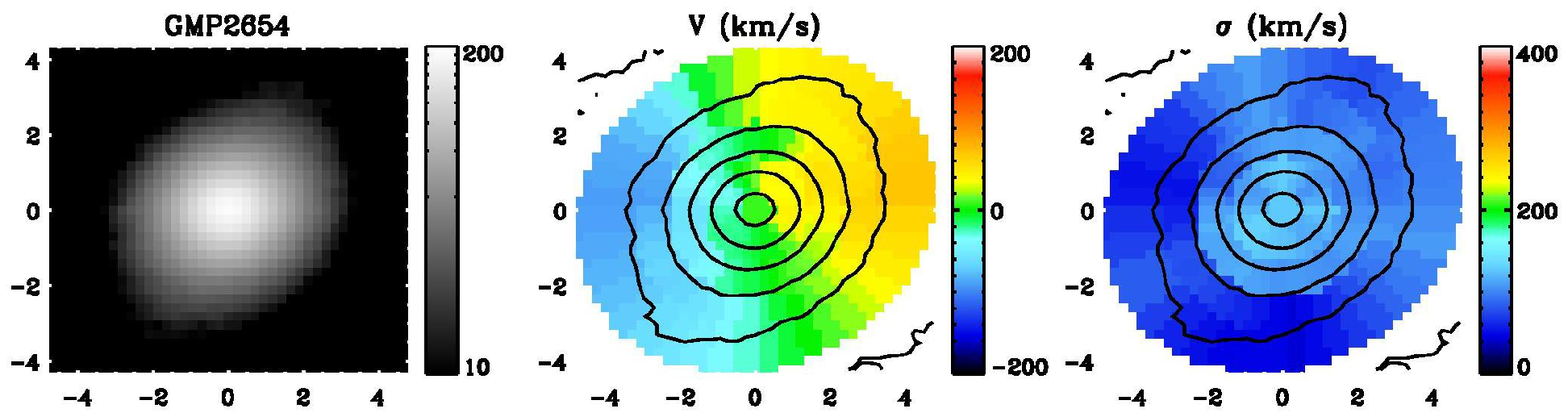}}
   
   \subfloat{\includegraphics[width=0.5\textwidth]{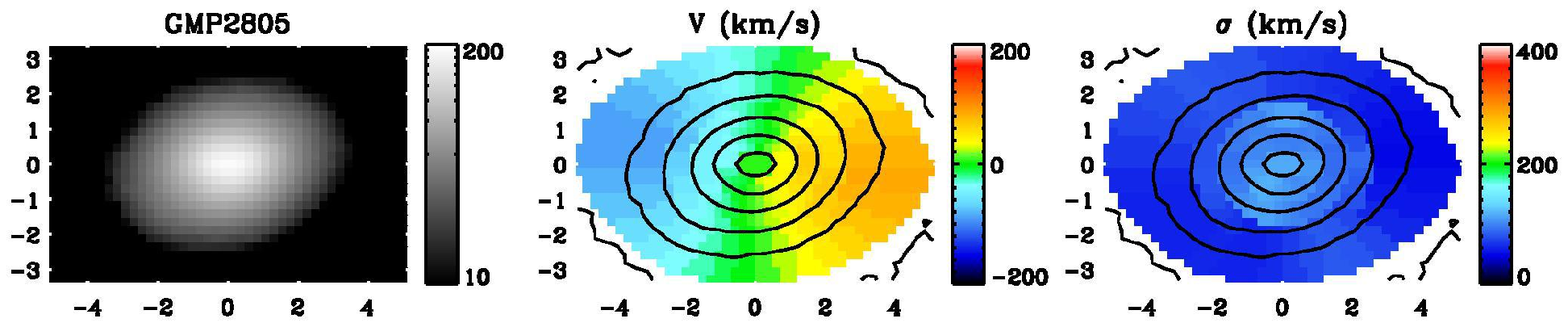}}      \subfloat{\includegraphics[width=0.5\textwidth]{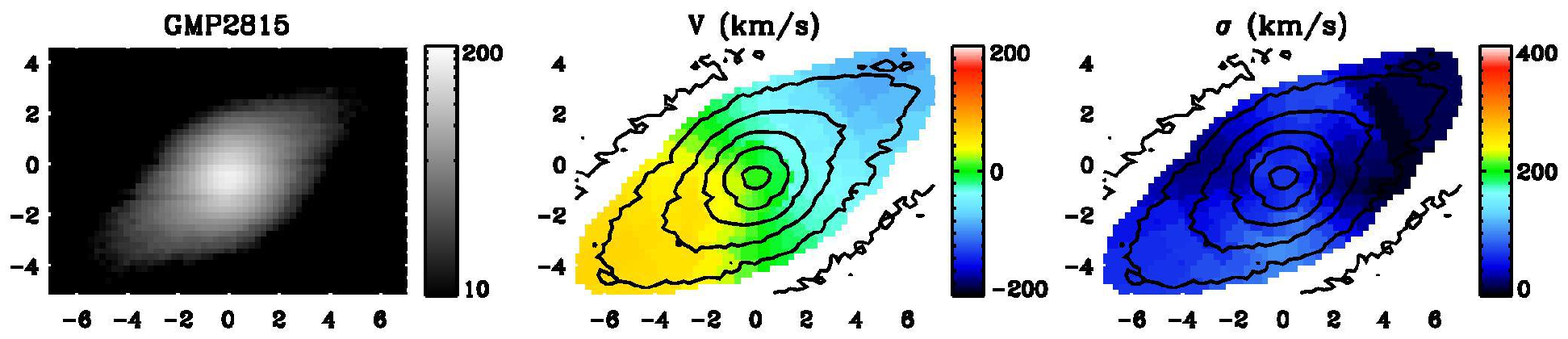}}
   
   \subfloat{\includegraphics[width=0.5\textwidth]{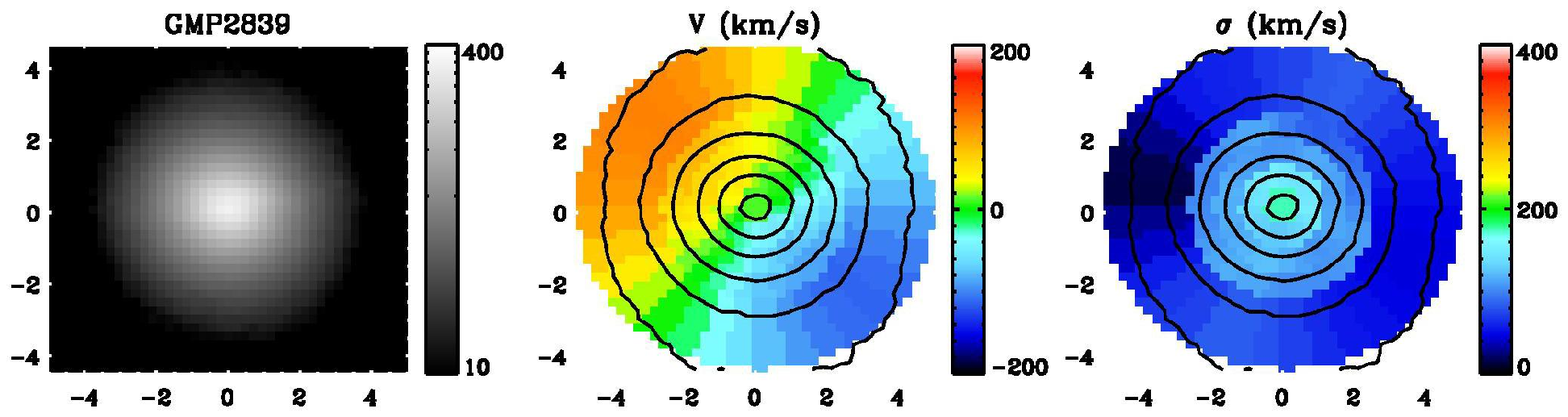}}      \subfloat{\includegraphics[width=0.5\textwidth]{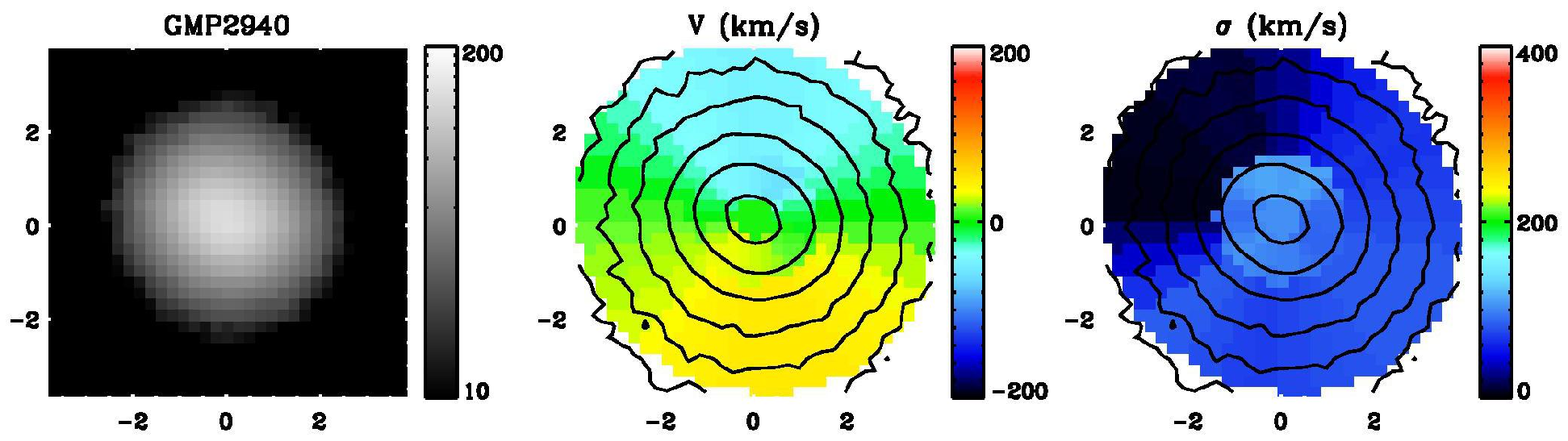}}
   
   \subfloat{\includegraphics[width=0.5\textwidth]{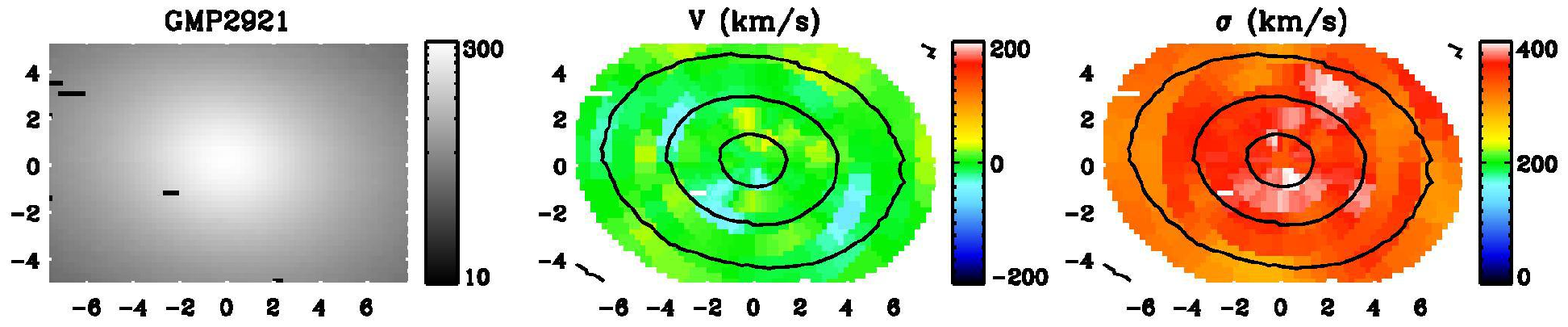}}      \subfloat{\includegraphics[width=0.5\textwidth]{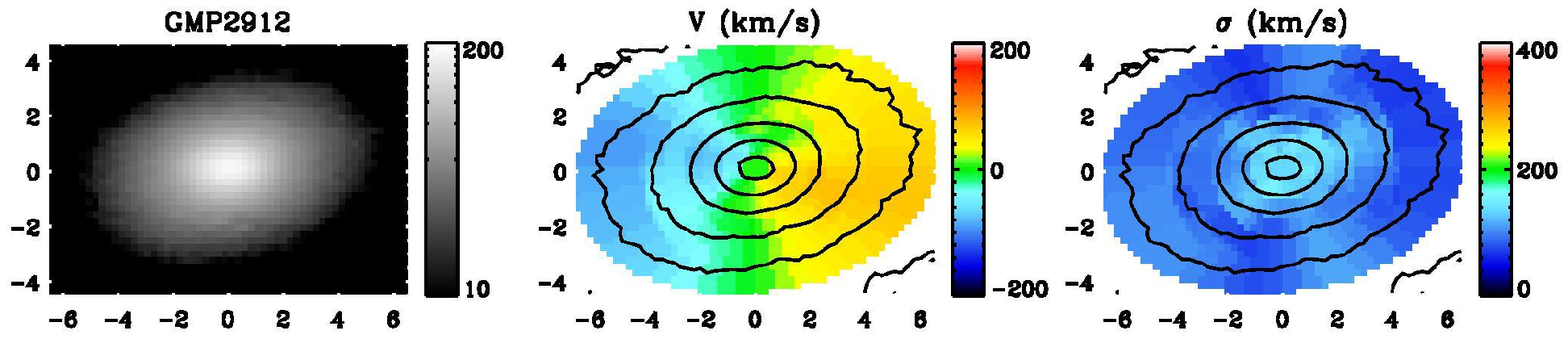}}
   
   \subfloat{\includegraphics[width=0.5\textwidth]{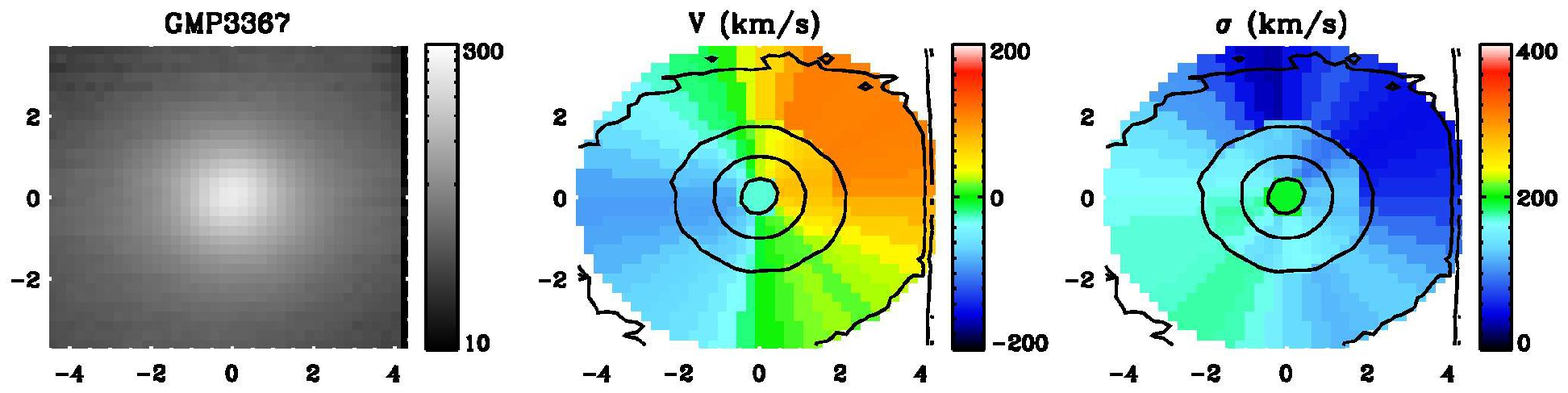}}      \subfloat{\includegraphics[width=0.5\textwidth]{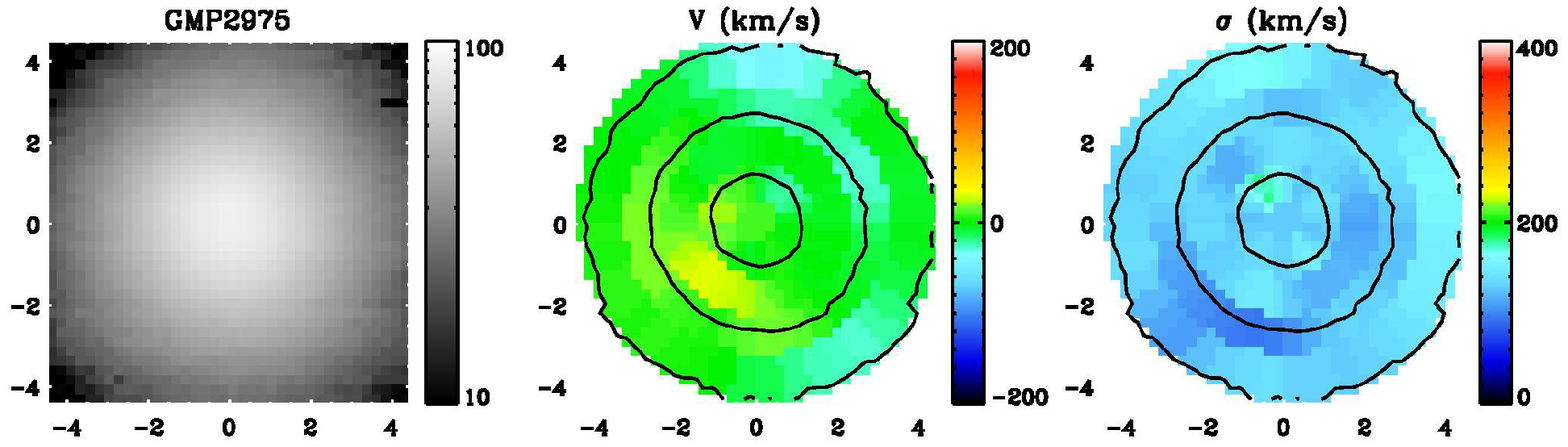}}
   
   \subfloat{\includegraphics[width=0.5\textwidth]{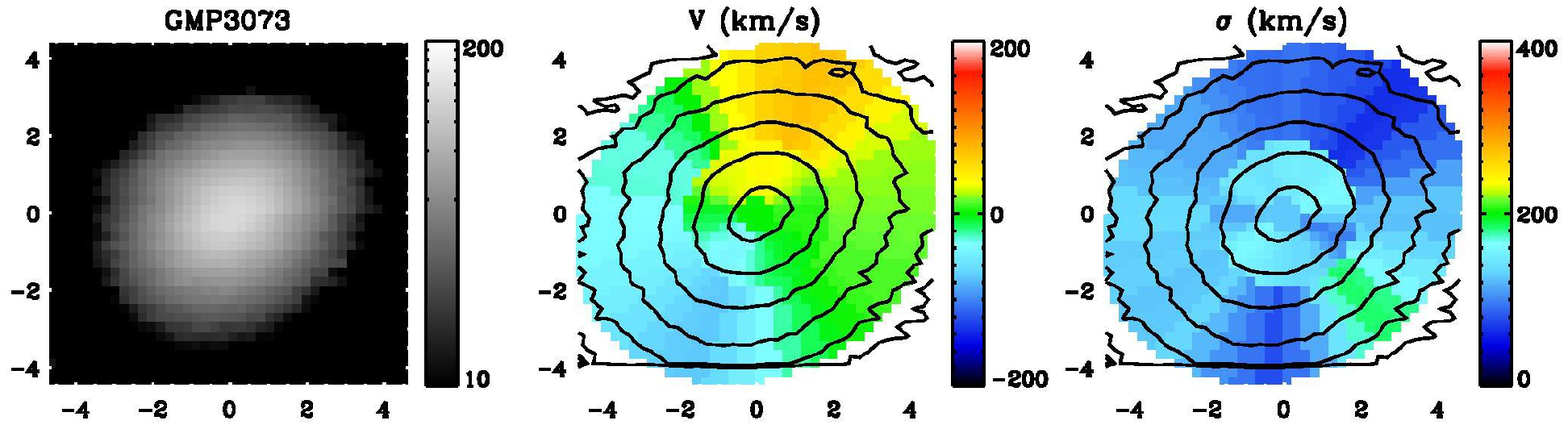}}      \subfloat{\includegraphics[width=0.5\textwidth]{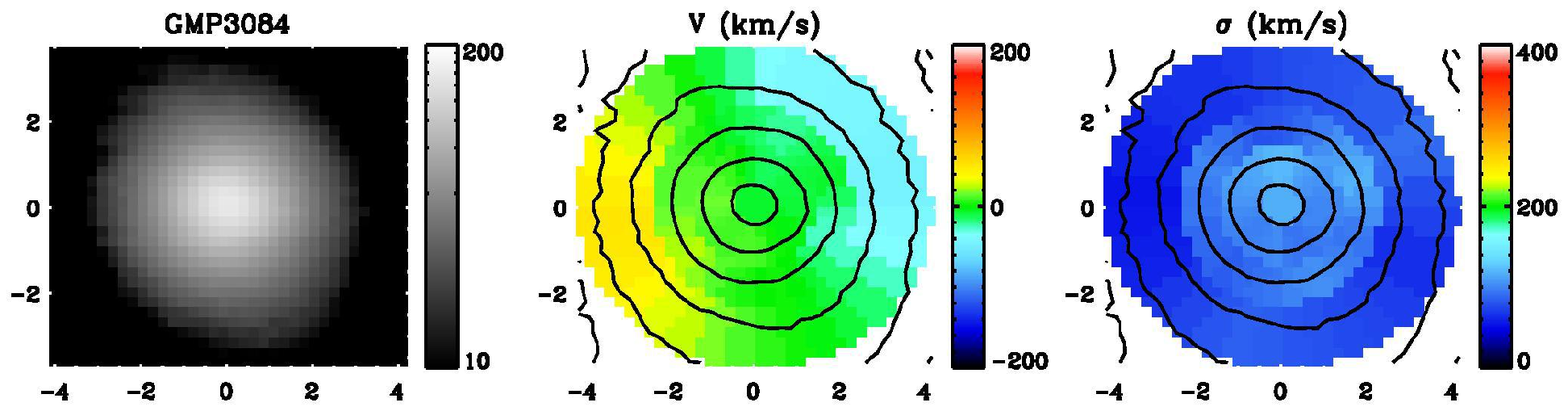}}
   
   \subfloat{\includegraphics[width=0.5\textwidth]{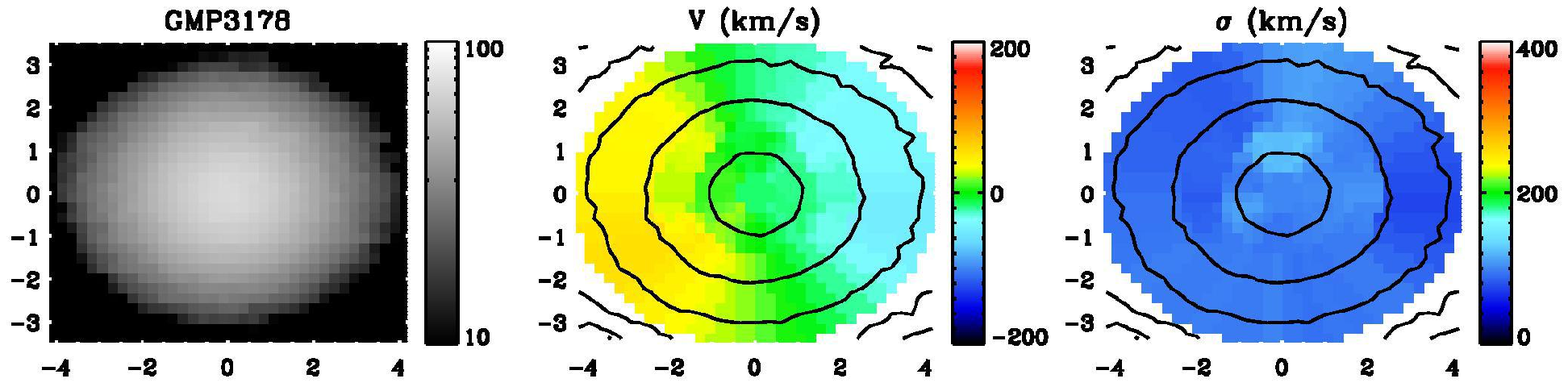}}      \subfloat{\includegraphics[width=0.5\textwidth]{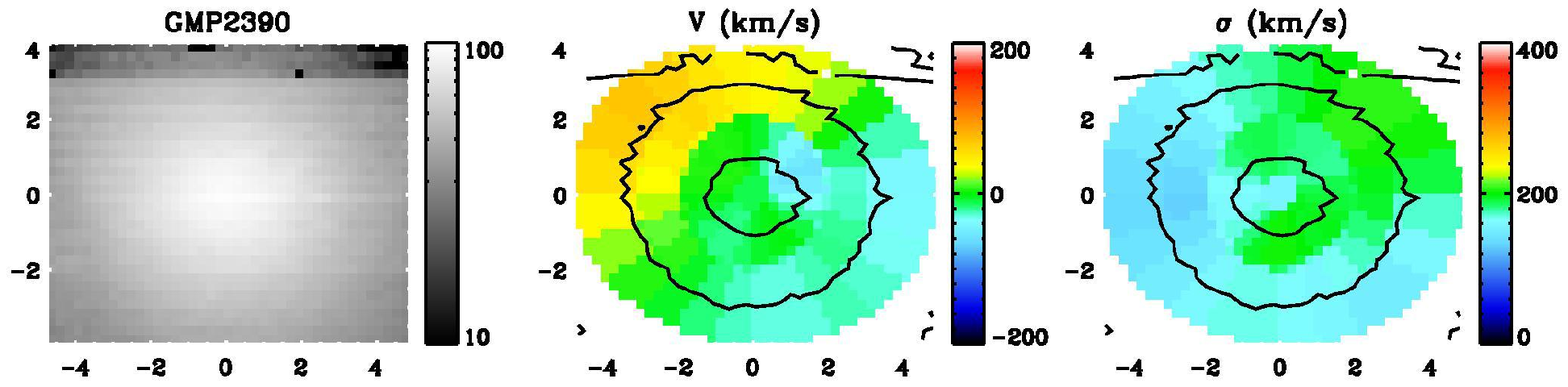}} 
   
   \caption{Continues on next page...}
\end{figure*}
\newpage
\begin{figure*}
   \ContinuedFloat
   \subfloat{\includegraphics[width=0.5\textwidth]{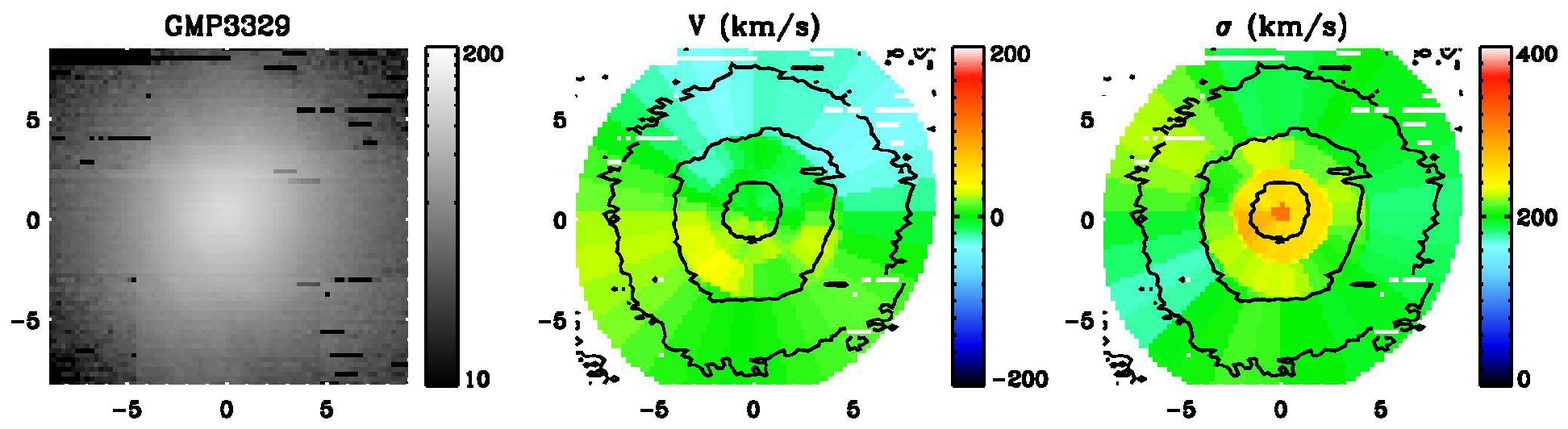}}      \subfloat{\includegraphics[width=0.5\textwidth]{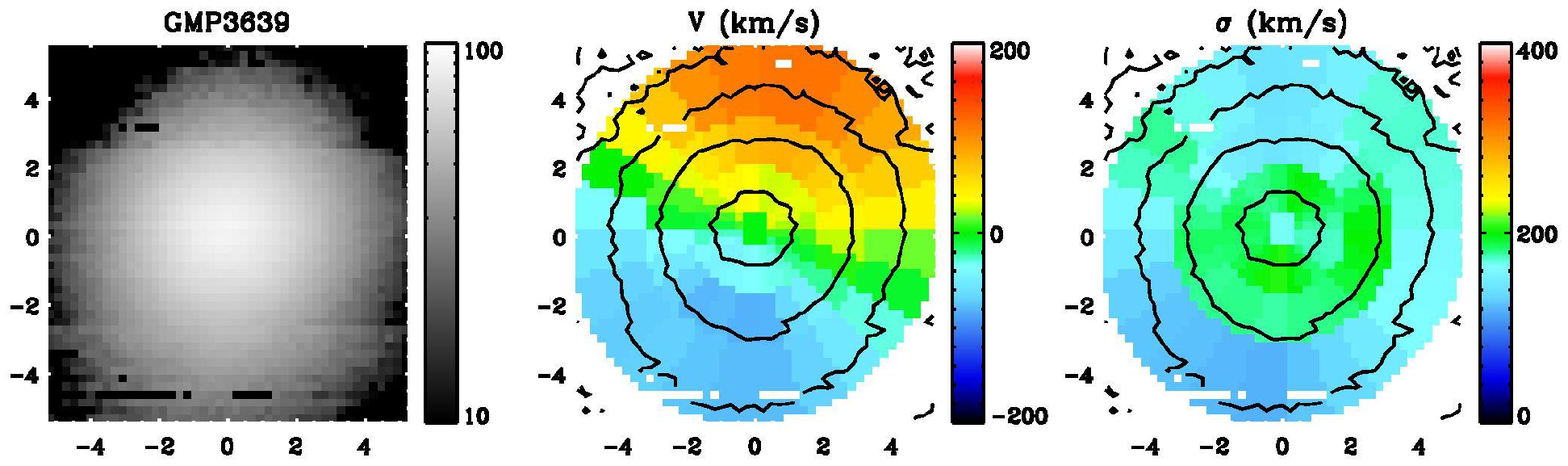}}
   
   \subfloat{\includegraphics[width=0.5\textwidth]{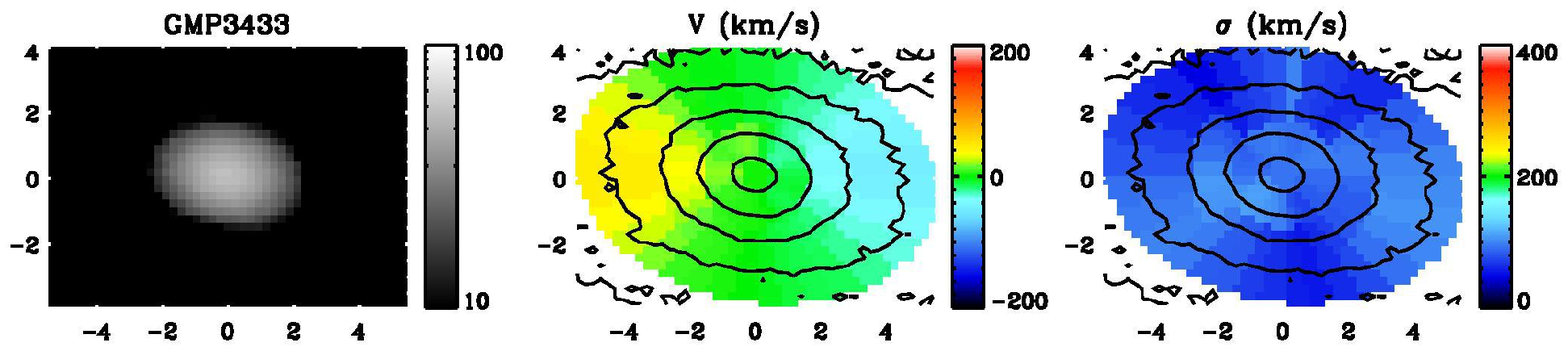}}      \subfloat{\includegraphics[width=0.5\textwidth]{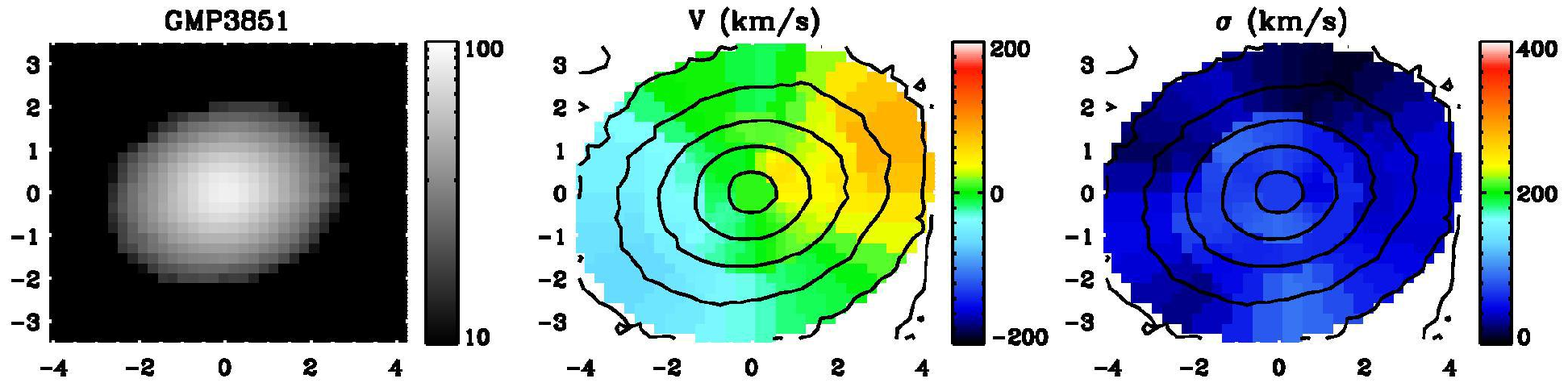}}
   
   \subfloat{\includegraphics[width=0.5\textwidth]{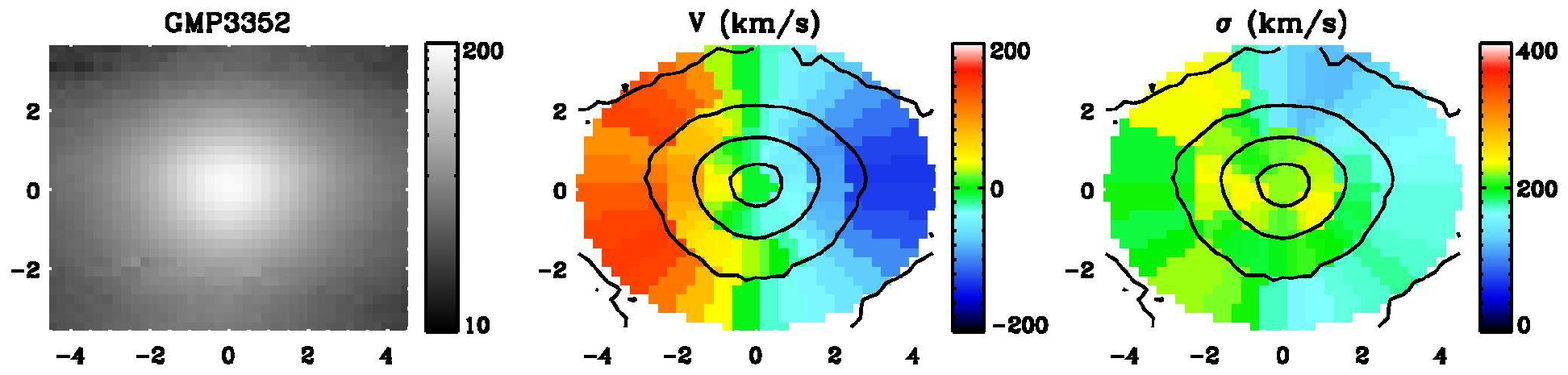}}      \subfloat{\includegraphics[width=0.5\textwidth]{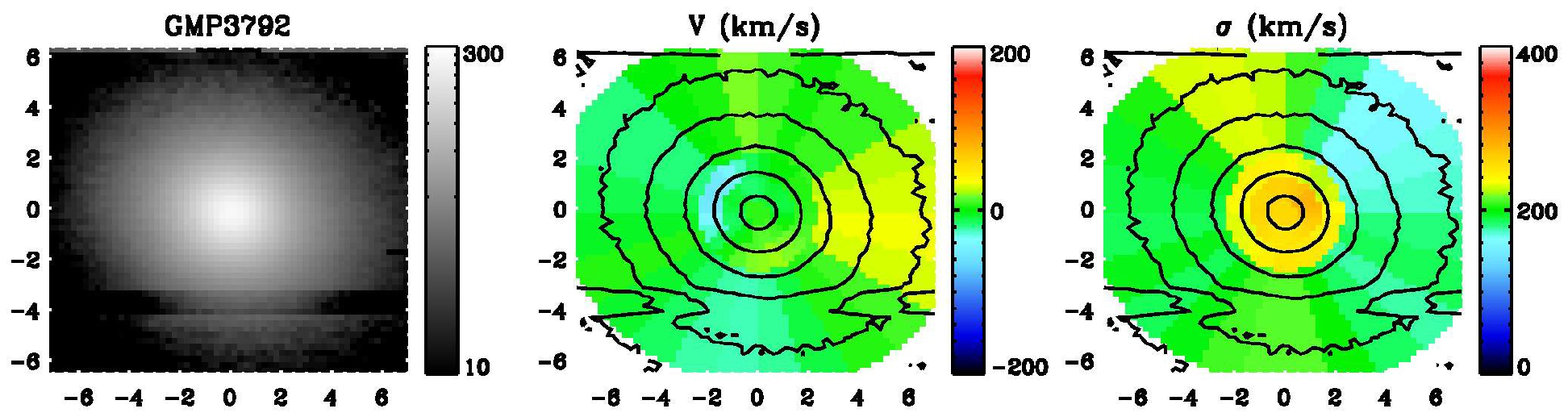}}
   
   \subfloat{\includegraphics[width=0.5\textwidth]{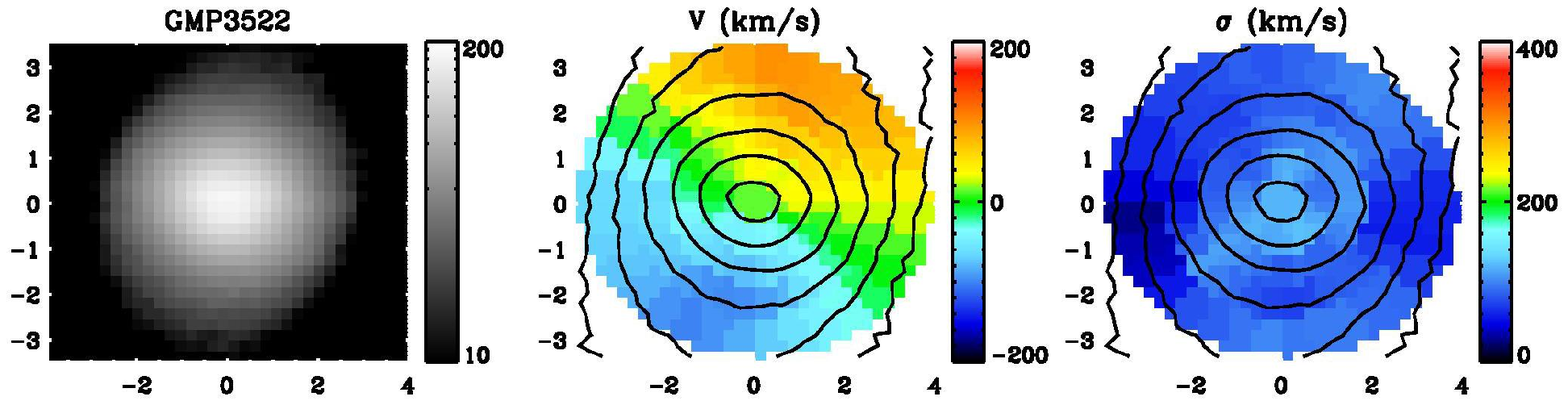}}      \subfloat{\includegraphics[width=0.5\textwidth]{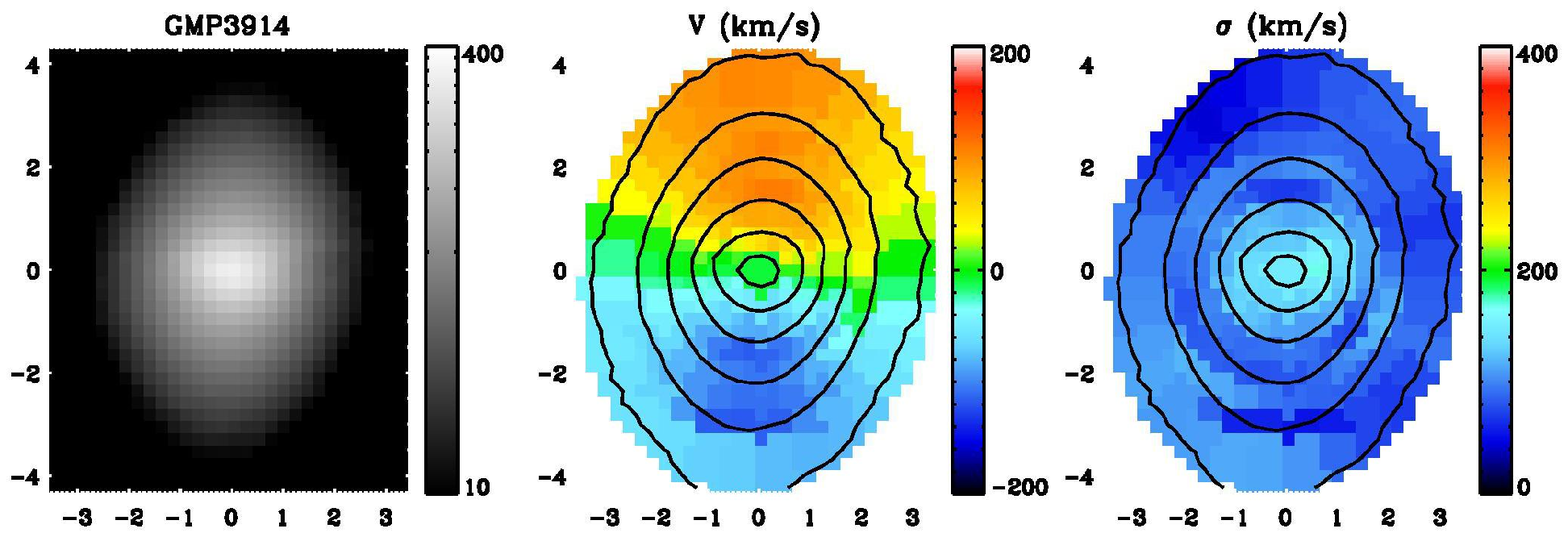}}
   
   \subfloat{\includegraphics[width=\textwidth]{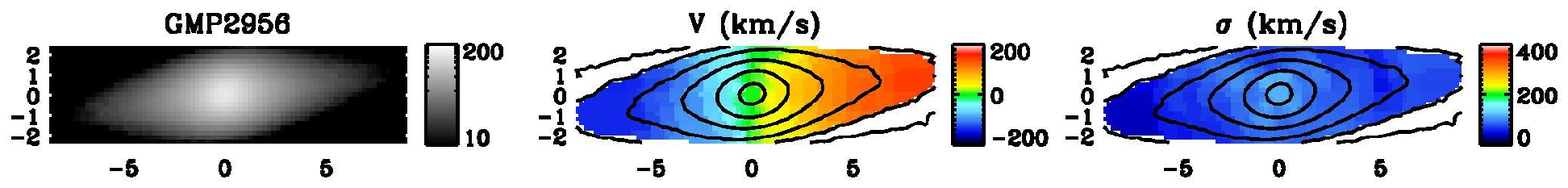}}
   
   \subfloat{\includegraphics[width=\textwidth]{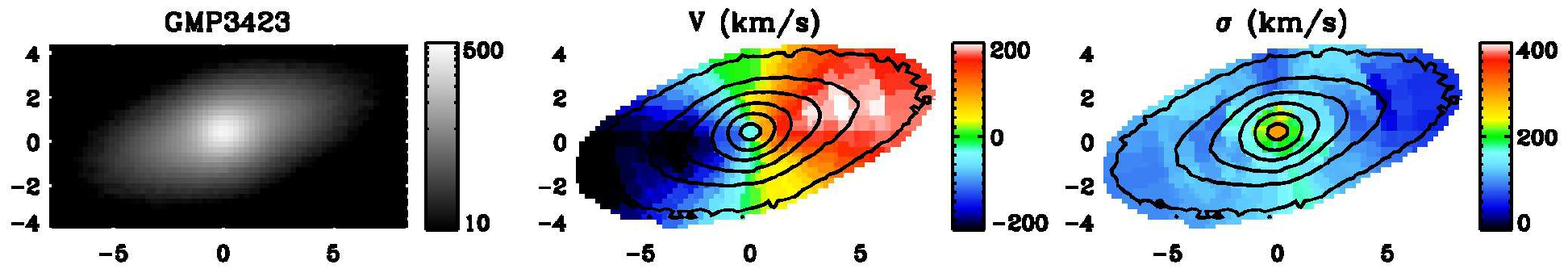}}
   
   \subfloat{\includegraphics[width=\textwidth]{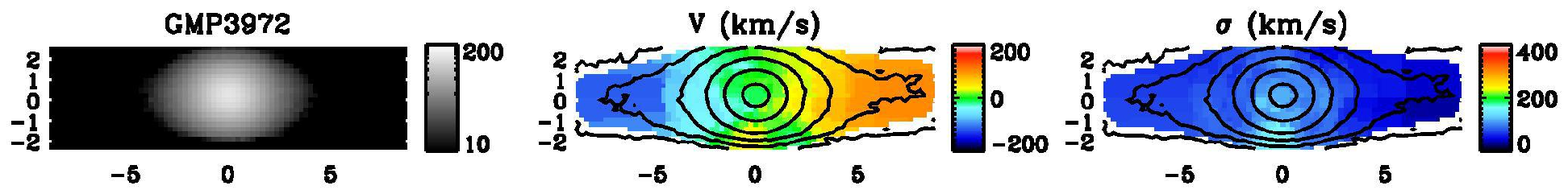}}
   \caption{SWIFT kinematics for \ngal early-type galaxies in the Coma cluster. For each galaxy, three images are shown: the flux map, the dithered velocity map and the dithered velocity dispersion map. X-axis and Y-axis scales are given in arc seconds. The ordering of the galaxies is not strictly sorted by the GMP I.D., but has been optimised to save space in the figure. The kinematic maps are averaged over many different binning realisations (see \S\ref{sec:data:kin}).} 
    \label{fig:kina}
\end{figure*}

\newpage

\section{Discussion}
\label{sec:discussion}

We presented a subset of our sample in \citep{Scott2012}. That subset did not sample the parent population fairly, but was biased to higher luminosities by design. \citet{Scott2012} found a higher \fsr than in this work, which is understood from the luminosity bias: SRs are more common than FRs for more more luminous galaxies. In the full sample presented here, we took precautions to be unbiased with respect to luminosity and ellipticity.

\begin{figure}
   \centering
   \includegraphics[width=0.5\textwidth]{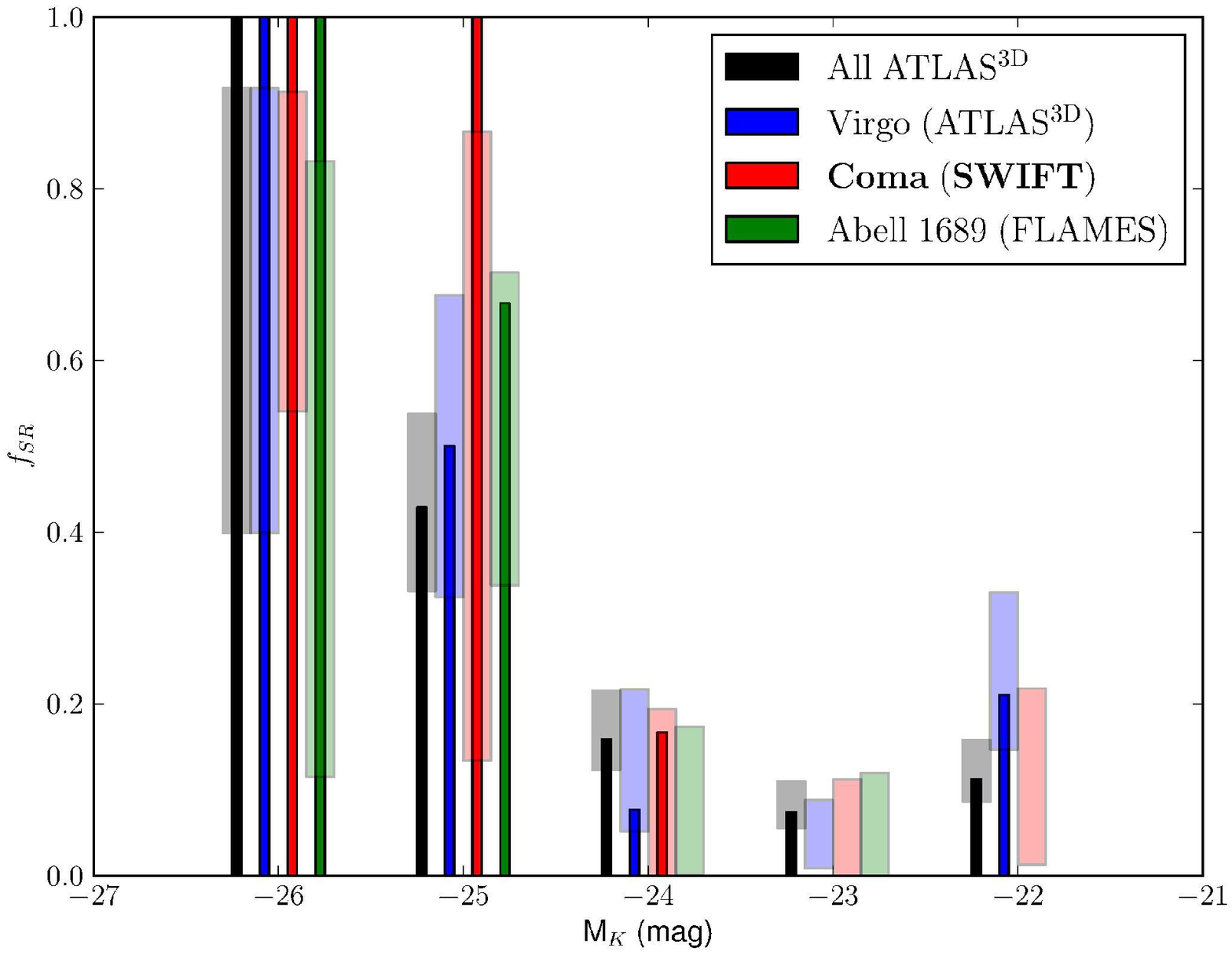} 
   \caption{The SR fraction (\fsr) in the ETG population as a function of absolute (2MASS) K magnitude: solid bars show the observed values while light shaded bars show the resulting uncertainty in the posterior. The entire \A3D sample is shown in black; Virgo data from \A3D is shown in blue; data for Abell~1689 is shown in green \citep{DEugenio2013} and the Coma data (this survey) is shown in red. The (posterior) uncertainties shown for the \A3D data assume a binomial distribution for \fsr, while the (posterior) uncertainties for Coma and Abell~1689 assume a hypergeometric distribution and account for measurement uncertainties (see \S\ref{sec:uncertainties}). }
   \label{fig:SRFRvMK}
\end{figure}

If we consider the dependence on luminosity, Fig. \ref{fig:SRFRvMK} confirms previous findings that more luminous galaxies are more likely to be SRs. Furthermore, this trend is no stronger or weaker for cluster galaxies than it is for the field/group galaxies in the \A3D survey, suggesting no dependence with GHE (except that the most luminous galaxies don't exist in the field/group environment of \A3D). This suggest that the formation mechanism for SRs must be more efficient for more luminous galaxies, but equally so across different GHEs. Although we find no low luminosity SRs, we sample enough low luminosity galaxies to place reasonable constraints on \fsr down to $M_\textrm{K}=-21.5$ mag and there is a suggestion of a non-zero number of SRs at the lowest luminosities from the calculation of our posterior $p(K | n, N)$. This originates from measurement uncertainty; a low luminosity FR (GMP3433) lies near the \lameps division and the uncertainties in \lam and \eps are not insignificant.

Combining our results for Coma with those of \citet{ATLAS3DVII} and \citet{DEugenio2013}, we find the \emph{average} \fsr in the ETG population is identical in the clusters studied so far (Virgo, Coma and Abell~1689). This is quite remarkable in itself, but even more remarkable is that this is the same as the average in the \A3D field and group environment. There appears to be no variation in \fsr with GHE. 

The morphology-density relation exists because ETGs are more prevalent in clusters, implying that the mechanism that produces ETGs is more efficient at higher (GHE) densities \citep{Dressler1980}. We have shown that \fsr is uniform across \emph{all} GHEs, which suggests that the mechanism that creates SRs shares exactly the same efficiency boost as the mechanism that produces FRs in a denser GHE. If the two mechanisms are one-and-the-same, this is a trivial consequence. But the structural differences between FRs and SRs strongly suggest different evolutionary histories and formation mechanisms. If one mechanism feeds off the success of the other (FRs might be transformed into SRs via merging), we might also expect to see a constant \fsr, so long as the mechanism efficiency (\fsr) is independent of the GHE density. Regardless, we now know that whatever formation mechanisms are proposed for FRs and SRs in the future, they must be independent of the GHE in which the ETGs are found today.

\begin{figure}
   \centering
   \includegraphics[width=0.5\textwidth]{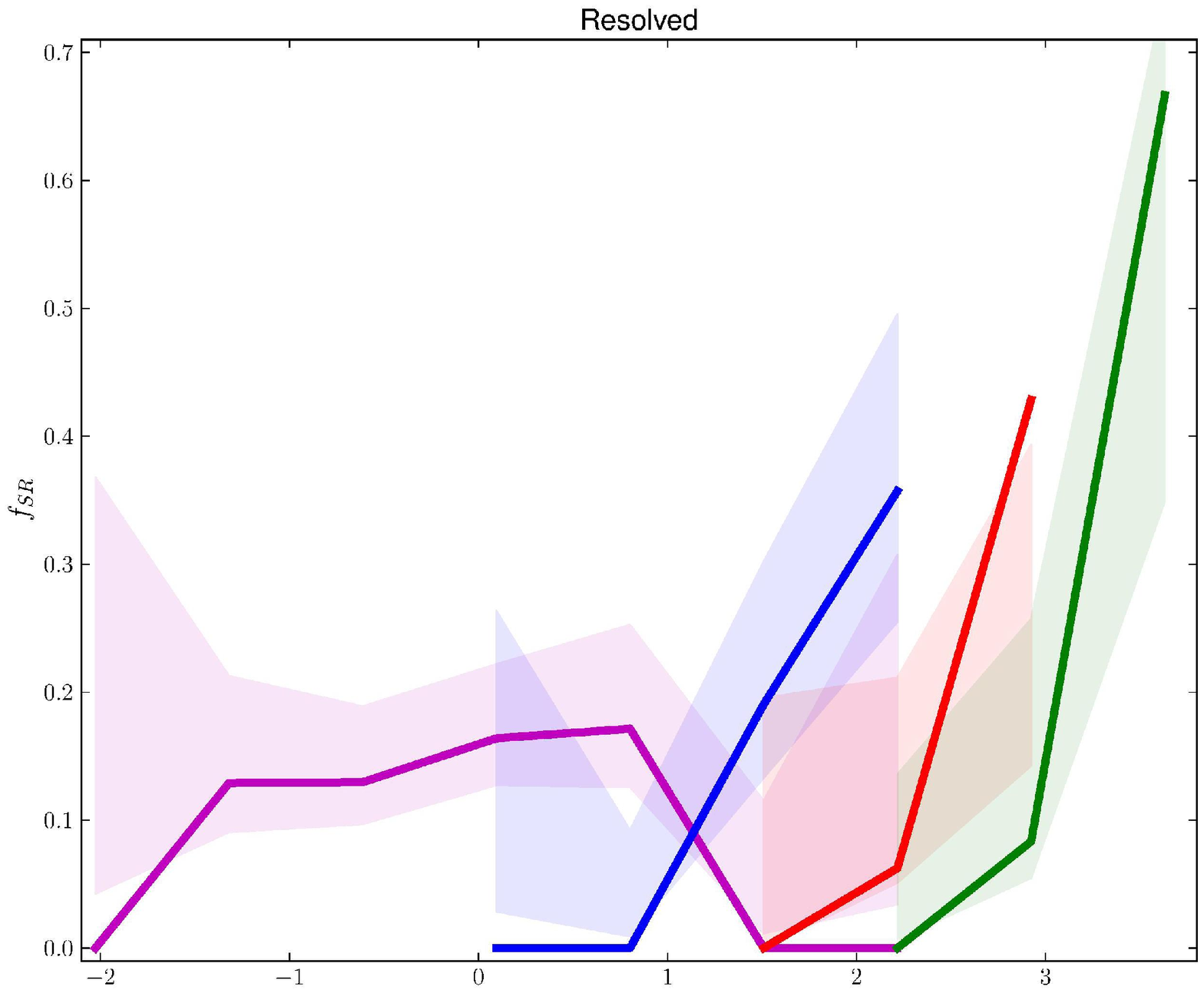} 
   \includegraphics[width=0.5\textwidth]{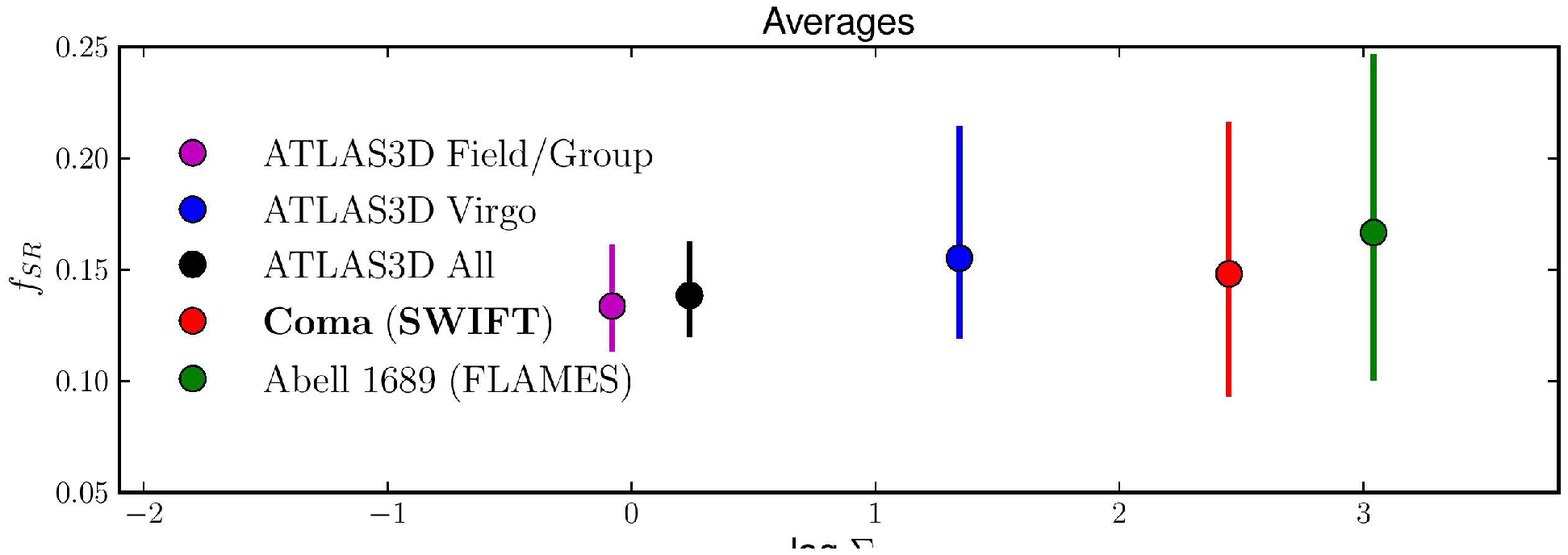} 
   \caption{The SR fraction (\fsr) in the ETG population as a function of projected density ($\log\Sigma_{3}$). \textbf{Upper}: resolving \fsr as a function of projected LPE density for each GHE; solid lines refer to the observed quantities, while shaded regions refer to the posterior uncertainties. \textbf{Lower}: the average \fsr for the average projected density of each GHE; for Coma and Abell~1689, average GHE densities refer to the mean LPE of the samples, not the mean LPE of the parent distribution. \textbf{All}: the entire \A3D volume limited survey is shown in black; \A3D data not in Virgo (field/group) is shown in magenta; \A3D data for Virgo alone is shown in blue; the data for Abell~1689 is shown in green \citep{DEugenio2013}; the data for Coma from this survey is shown in red. The uncertainties shown for the \A3D data assume a binomial distribution for \fsr, while the uncertainties for Coma and Abell~1689 assume a hypergeometric distribution and account for measurement uncertainties (see \S\ref{sec:uncertainties}).}
   \label{fig:SRFRvSIG}
\end{figure}

More insights can be gained from resolving clusters into subpopulations of different LPE density (Fig. \ref{fig:SRFRvSIG}). For Coma and Abell~1689, we see strong evidence that SRs are more likely to be present in the denser LPEs within a cluster. This is not a threshold effect: there is no special number density that catalyses the formation of SRs. On the contrary, we see that galaxies in Abell~1689 with $\log\Sigma_{3} \approx 2.2$ are unlikely to be SRs (below the cluster average), while galaxies at the same densities in Coma and Virgo have a higher chance of being SRs (above the cluster average). Given that these clusters share the same average \fsr, there must be a process that segregates SRs from the overall population, forcing them towards the densest regions of the cluster. As discussed in \citet{DEugenio2013}, such a process could be dynamical friction. We know that SRs are far more likely to be the most massive galaxies in the cluster, and that dynamical friction is more effective for massive galaxies. One concern for this proposal is that SRs could then merge, depleting their numbers; this may \emph{decrease} \fsr in dense LPEs, unless there is a mechanism to prevent this merging, such as in the \emph{over merging} problem \citep{Moore1999}. Perhaps FRs are transformed into SRs at the same rate that SRs merge with themselves; such delicate balance permits a uniform \fsr across different GHEs while also allowing for the LPE gradient that we see in individual clusters. Such a scenario is almost hierarchical and could explain why massive galaxies are more likely to be SRs; it also suggests that the most massive galaxies in any given GHE will be SRs.

This study and that of \citet{DEugenio2013} both find that the region of \lam populated by FRs in clusters is different to the region populated by the \A3D sample; FRs in Coma and Abell~1689 generally have $\lambda<0.6$ whereas the \A3D sample populates $\lambda<0.8$. Taken at face value, this suggests a different distribution for the intrinsic anisotropy of the galaxies: \citet{SAURONX} and \citet{ATLAS3DIII} found that the population of FRs in the \lameps diagram could be explained by projection of anisotropic rotators, with the anisotropy distributed Normally but truncated to exclude anisotropies $\delta > 0.8 \epsilon_\textrm{int}$, where $\epsilon_\textrm{int}$ is the intrinsic ellipticity (i.e. not projected). The effect of truncating the anisotropy distribution is to exclude $\lambda>0.8$. If one were to truncate at a lower anisotropy, one could similarly exclude lower values of \lam. This would suggest there is a mechanism in the cluster environment that lowers the anisotropy of FRs. Further studies with much larger sample sizes could better investigate changes in the distribution of FRs in the \lameps plane. Similarly, studies of the observed ellipticity of galaxies could corroborate these claims. 

Finally we note that the dispersion maps can show strong, weak and sometimes no central dynamically hot component. These components are assumed to originate from a stellar bulge, which may be strong, weak or absent. Such \emph{kinematic} identification of bulges may help with decomposing these galaxies into disk and bulge components, or similarly may help prevent unnecessary parametrisation (and degeneracy) in the surface photometry when no kinematic bulge is evident. The lack of a hot central component in some galaxies suggests that bulges are not ubiquitous in ETGs.

\subsection{Future Studies}

To conclusively understand the processes that create SRs, more investigation is required. Our limited sample of \ngal galaxies has barely sampled the entirety of Coma. The Coma cluster is known to extend over 3 degrees but we only sample galaxies within a radius of 15\arcmin (0.43 Mpc) from the cluster centre. Similarly, the study of Abell~1689 by \citet{DEugenio2013} was limited to the HST/ACS footprint (202\arcsec$\times$200\arcsec), which covers a (circularised) radius of only 0.35 Mpc. By comparison, a radius of 0.5 Mpc in Virgo enclosed 8 of the 9 SRs \citep{ATLAS3DVII}; clearly future studies of the cluster outskirts are required to establish if only FRs are found outside a critical radius, which would tend to \emph{lower} the average \fsr found for both Coma and Abell~1689. Furthermore, we have poor statistics on \fsr at lower LPE density: the presence of SRs in the low density outskirts of clusters may challenge the need for dynamical friction. At present, although our measurements for Coma show a gradient in \fsr with \sig3, the posterior uncertainties are consistent with no gradient. To improve on this result requires a much larger survey; for example, were a complete IFS survey of Coma's central square arcminute ($\sim150$ galaxies) to find the same gradient as seen in Fig. \ref{fig:SRFRvSIG}, the posterior uncertainties (assuming the same binning, negligible measurement uncertainty and a binomial distribution for \fsr) would not overlap at the 68\% CI level. Such a level of accuracy would provide meaningful constraints on future model predictions. Similarly, $kT-\Sigma$ data only exists for three clusters, which is far from a representative sample, and significantly lacking compared to the 55 clusters studied by \citet{Dressler1980}. The local Universe contains many different groups and clusters (GHEs) with a range of properties which remain to be surveyed with IFS. We also have poor understanding of low luminosity SRs, which appear to be rare. A study concentrating on intermediate GHE densities, such as groups, could help us understand if a true hierarchy exists, and if the most massive galaxies in a local volume or parent halo (GHE) always have low specific angular momentum. New multiplexed IFS instruments like SAMI \citep{Croom2012,Fogarty2013}, KMOS \citep{KMOS} and MaNGA have the potential to perform these studies efficiently. A more targeted approach could study individual massive galaxies at higher redshifts in an attempt to witness the assembly of SRs, be it a sudden or gradual process. 

Our theoretical understanding of the processes that create SRs is currently undergoing change. Until recently, it was widely accepted that pressure supported, slowly rotating systems were the end products of major mergers \citep{Barnes1988,Hernquist1992,NaabBurkert2003}. But the latest N-body simulations of binary mergers with realistic, cosmologically motivated impact parameters suggest that the orbital angular momentum becomes locked in the stars of the remnant, leading to significant specific angular momentum and flattening in all but the most elaborate (and unlikely) initial configurations \citep{Bois2011}. Given the segregation of SRs into regions of higher LPE density suggests a dominant role for dark matter, it may be revealing to run similar models inside group- or cluster-scale haloes with increased dynamical friction; we know for example that the two central SRs in Coma will eventually merge \citep{Gerhard2007}. Semi-analytic models can produce SRs if the most massive galaxies in massive halos are allowed to cannibalise material from tidally stripped satellites \citep{ATLAS3DVIII}. These studies now need to turn to group-like environments with lower (GHE) densities to identify if the same mechanisms can maintain a constant \fsr there. Furthermore, studying individual clusters in these models may explain the excess of SRs towards the densest LPEs, and the absence of SRs in the lowest density LPEs, while the average \fsr remains constant across different GHE densities. 

\section{Conclusion}
\label{sec:conclusion}

Using the Oxford SWIFT spectrograph, we have surveyed \ngal galaxies in the Coma cluster, taking care to minimise sample bias with respect to luminosity and ellipticity. We find \NSR slow rotators, all of which have \MK$<24$~mag and $\epsilon<0.4$. The average slow rotator fraction in the Coma ETG population is thus \SRfrac. This is identical to the average SR fraction found in the \A3D field/group environment as well as the Virgo and Abell~1689 clusters, suggesting no change with GHE. However, within the clusters the distribution of slow rotators is not uniform, but appears to be concentrated towards denser LPEs. We confirm that the SR fraction is higher at higher luminosities, and find no variation of the distribution with GHE. These results constrain the different mechanisms needed to produce the contrasting physical properties of fast and slow rotators. Both mechanisms must increase in efficiency in clusters to produce the excess of ETGs observed, while at the same time maintaining a constant SR fraction inside each GHE. Conversely, the mechanism for producing FRs is more efficient at lower luminosities, but no more so across different GHEs. 

We also find that the velocity dispersion maps of FRs are generally dynamically colder than the dispersion maps of SRs. Furthermore, dynamically hot central components (presumably relating to the presence of a bulge) can be seen at the centres of \emph{some}, but not all, FRs.

\section*{Acknowledgments}
We thank the anonymous referee for their excellent comments and suggestions, which significantly improved this work. We also thank the staff at the Palomar Observatory for their help and support with the observations.
RCWH was supported by the Science and Technology Facilities Council
[STFC grant number ST/H002456/1].
FDE acknowledges support from the Physics Department, University of Oxford and travel support from Merton College, Oxford.
RLD acknowledges travel and computer grants from Christ Church College, Oxford.
NS acknowledges support of Australian Research Council grant DP110103509.
The Oxford SWIFT spectrograph is directly supported by a Marie Curie Excellence Grant from the European Commission (MEXT-CT-2003-002792, Team Leader:
N. Thatte). It is also supported by additional funds from
the University of Oxford Physics Department and the John
Fell OUP Research Fund. Additional funds to host and support SWIFT at the 200-inch Hale Telescope on Palomar are
provided by Caltech Optical Observatories. This research is based on observations obtained at the Hale Telescope,
Palomar Observatory, as part of a collaborative agreement
between the California Institute of Technology, its divisions
Caltech Optical Observatories and the Jet Propulsion Laboratory (operated for NASA), and Cornell University.
Funding for the SDSS and SDSS-II has been provided by the Alfred P. Sloan Foundation, the Participating Institutions, the National Science Foundation, the U.S. Department of Energy, the National Aeronautics and Space Administration, the Japanese Monbukagakusho, the Max Planck Society, and the Higher Education Funding Council for England. The SDSS Web Site is http://www.sdss.org/.
The SDSS is managed by the Astrophysical Research Consortium for the Participating Institutions. The Participating Institutions are the American Museum of Natural History, Astrophysical Institute Potsdam, University of Basel, University of Cambridge, Case Western Reserve University, University of Chicago, Drexel University, Fermilab, the Institute for Advanced Study, the Japan Participation Group, Johns Hopkins University, the Joint Institute for Nuclear Astrophysics, the Kavli Institute for Particle Astrophysics and Cosmology, the Korean Scientist Group, the Chinese Academy of Sciences (LAMOST), Los Alamos National Laboratory, the Max-Planck-Institute for Astronomy (MPIA), the Max-Planck-Institute for Astrophysics (MPA), New Mexico State University, Ohio State University, University of Pittsburgh, University of Portsmouth, Princeton University, the United States Naval Observatory, and the University of Washington.
This research made use of Montage, funded by the National Aeronautics and Space Administration's Earth Science Technology Office, Computational Technnologies Project, under Cooperative Agreement Number NCC5-626 between NASA and the California Institute of Technology. The code is maintained by the NASA/IPAC Infrared Science Archive.
This research has made use of the NASA/IPAC Extragalactic Database (NED) which is operated by the Jet Propulsion Laboratory, California Institute of Technology, under contract with the National Aeronautics and Space Administration.
\appendix

\section{Sources of uncertainty in \lam}
\label{sec:appendix:lamerrs}
Here we list dominant sources of uncertainty and our approaches to quantifying and, where possible, minimising their contribution in the calculation of \lam.

\subsection{Random uncertainty from photon shot noise}
\label{sec:appendix:randomerror}

As explained in \S\ref{sec:uncertainties}, the formal random uncertainties in \V and \s provided by \ppxf originate directly from the photon noise. However, the propagation of these uncertainties through the expression for the calculation of \lam from \V and \s is not trivial and is not documented. We summarise the formulae for propagation here, assuming a standard first order (Taylor expansion) approach. 

A standard first order Taylor expansion tells us that if the uncertainties are small, we can propagate random uncertainties from input parameters using the first derivatives with respect to those parameters. Thus for \lam, considering only covariances between \V and \s,
\begin{eqnarray}
(\Delta \lambda)^{2} & \approx & \left( \frac{\pd \lambda}{\pd F_{i}}  \Delta F_{i}\right)^{2} + \left( \frac{\pd \lambda}{\pd R_{i}} \Delta R_{i}\right)^{2} \\
& + & \left( \frac{\pd \lambda}{\pd V_{i}} \Delta V_{i}\right)^{2} + \left( \frac{\pd \lambda}{\pd \sigma_{i}} \Delta \sigma_{i}\right)^{2}\\
& + & 2 \frac{\pd \lambda}{\pd V_{i}}\frac{\pd \lambda}{\pd \sigma_{i}}\textrm{Cov}(V_{i}, \sigma_{i}).
\end{eqnarray}
The first order derivatives with respect to each input parameter are
\begin{equation}
\frac{\pd \lambda}{\pd F_i} = \frac{R_{i} |V_{i}|}{\displaystyle\sum\limits_{j}F_{j}R_{j}M_{j}} - R_{i} M_{i}  \left[\frac{\displaystyle\sum\limits_{k}F_{k} R_{k} |V_{k}|}{\left(\displaystyle\sum\limits_{l}F_{l} R_{l} M_{l}\right)^{2}}\right]  
\end{equation}
\begin{equation}
\frac{\pd \lambda}{\pd R_i} = \frac{F_{i} |V_{i}|}{\displaystyle\sum\limits_{j}F_{j}R_{j}M_{j}} - F_{i} M_{i}  \left[\frac{\displaystyle\sum\limits_{k}F_{k} R_{k} |V_{k}|}{\left(\displaystyle\sum\limits_{l}F_{l} R_{l} M_{l}\right)^{2}}\right] 
\end{equation}
\begin{equation}
\frac{\pd \lambda}{\pd V_i} = \frac{F_{i} R_{i} \textrm{Sgn}(V_{i})}{\displaystyle\sum\limits_{j}F_{j}R_{j}M_{j}} - \frac{F_{i} R_{i}|V_{i}|}{M_{i}}  \left[\frac{\displaystyle\sum\limits_{k}F_{k} R_{k} |V_{k}|}{\left(\displaystyle\sum\limits_{l}F_{l} R_{l} M_{l}\right)^{2}}\right] 
\end{equation}
\begin{equation}
\frac{\pd \lambda}{\pd \sigma_i} = - \frac{F_{i} R_{i} \sigma_{i}}{ M_{i}}  \left[\frac{\displaystyle\sum\limits_{k}F_{k} R_{k} |V_{k}|}{\left(\displaystyle\sum\limits_{l}F_{l} R_{l} M_{l}\right)^{2}}\right] 
\end{equation}
where $M_{i} = (V_{i}^{2}+\sigma_{i}^{2})^{1/2}$ and $\textrm{Sgn}(V_{i})=V_{i}/|V_{i}|$. The choices for $\Delta V_{i}$ and $\Delta \sigma_{i}$ are obvious and come straight from the standard errors provided by \ppxf; $\Delta F_{i}$ originates from $\sqrt{N}$ where $N$ is the number of photons collected in that bin while $\Delta R_{i}$ was chosen as the standard deviation of the radii of the spaxels in each bin (thus bins extending over larger radii have a larger $\Delta R_{i}$). In practice, the uncertainties in \V and \s dominate and the typical (mean) relative uncertainty in \lam is 10\%. 

\subsection{Discretisation noise}
\label{sec:appendix:binningerror}

It became apparent that the discretisation of the velocity and velocity dispersion maps (an unavoidable consequence of binning to the S/N necessary for extraction of kinematics) leads to a source of noise in \lam; different choices in the sizes and positions of the bins lead to different kinematics and measures of \lam. We wish to quantify this effect in order to establish its significance. This is not trivial as most binning mechanisms are not random (and so cannot be seeded to give different configurations) but generate a single unique configuration. However, in the sector approach we use here, the azimuthal angle used for each sector can be rotated by a fraction of the sector width, leading to slightly different binning choices. Applying different rotations, we generated many different configurations; comparing the kinematics extracted in each case allowed us to estimate the noise generated by discretisation. This approach suggests that the relative uncertainty in \lam from discretisation of the \V and \s maps is around 5\% which is half the typical uncertainty from photon noise and can further be reduced by use of \emph{dithered} kinematic maps (\S\ref{sec:data:kin}).  It is therefore not a significant source of uncertainty. 

\subsection{Systematic uncertainty from sky line subtraction residuals}
\label{sec:appendix:skyerror}

We found that residuals from the subtraction of sky emission lines can cause a catastrophic failure in the kinematic results (the model fit not representing the underlying galaxy spectrum), leading to error in the measurement of \V and \s and thus \lam. Such systematics can be significantly reduced, if not removed completely, by masking strong sky lines, or simultaneously fitting the sky spectrum when fitting the kinematics in \ppxf. However, masking can also remove one or more of the calcium triplet absorption lines from the fit (i.e. information on \V and \s), so it is undesirable to always mask the sky emission lines; conversely, fitting the sky spectrum simultaneously retains all the absorption lines but also retains residuals from imperfect subtraction. To avoid ad-hoc decisions on when to mask or simultaneously fit the sky, it is desirable to define physically motivated reasons for a systematic approach in all cases. 

\begin{figure}
   \centering
   \includegraphics[width=0.45\textwidth]{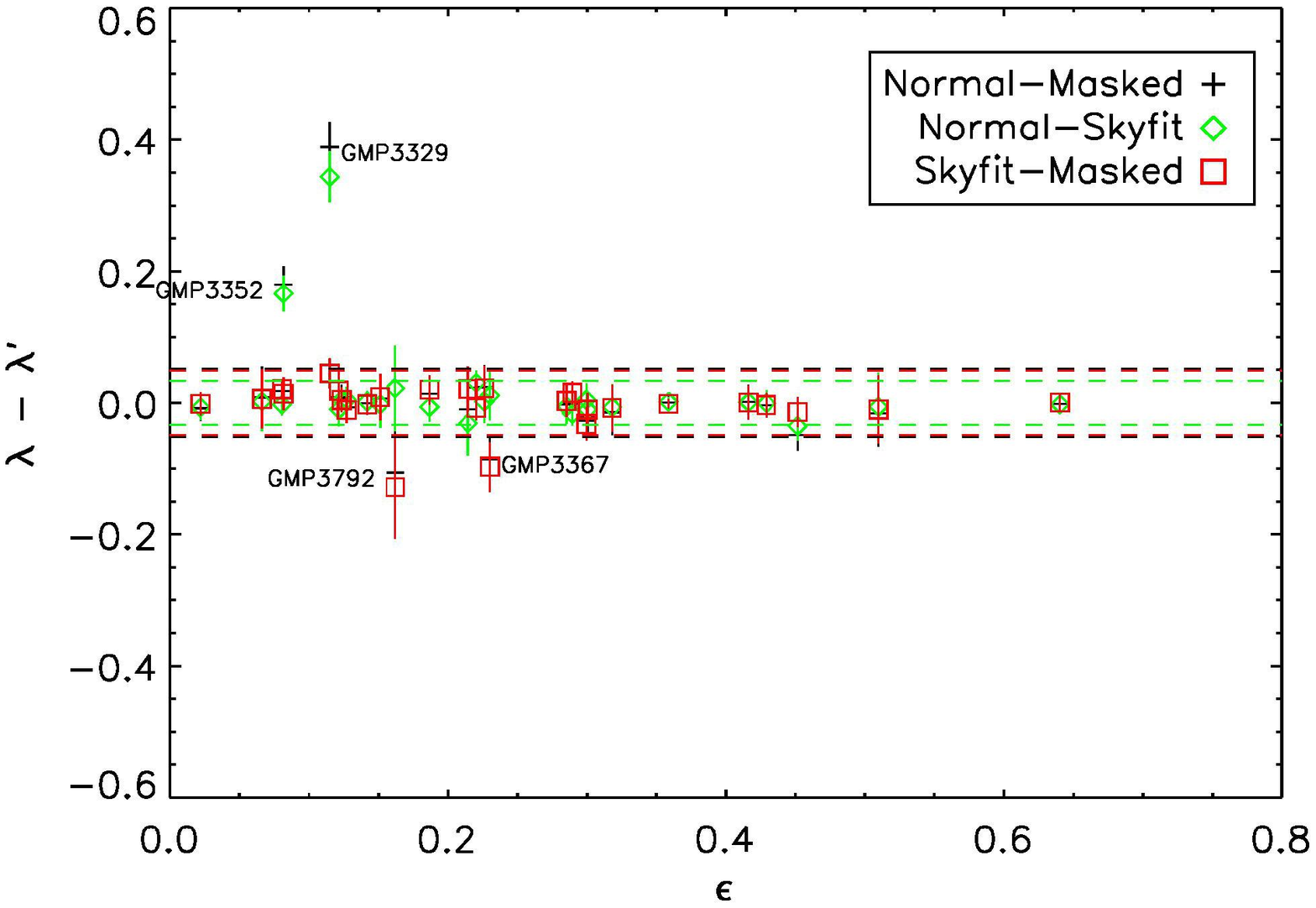} 
   \caption{The difference in the calculation of \lam with/without masking and simultaneous fitting of the sky spectrum. The difference in \lam with and without masking is shown in black; the difference with and without simultaneous sky spectrum fitting is shown in green and the difference between masking and simultaneous fitting is shown in red. For each case, a robust estimation of the standard deviation is shown by the dashed lines.}
   \label{fig:appendix:lamdiff}
\end{figure}

We extracted the kinematics and measured \lam (from dithered maps) with and without masking of sky emission lines and compare the results in Fig. \ref{fig:appendix:lamdiff} (black). Sky residuals do not bias the determination of \lam in the majority of cases, but the few cases where there is an error need to be corrected. A robust estimation of the standard deviation (using the IDL {\sc robust\_sigma} algorithm) gives $\sigma_{\lambda-\lambda^{\prime}}=0.017$.

As an alternative to masking the affected regions, we can fit the sky spectrum simultaneously while fitting the kinematics \citep[as done in][]{Weijmans2009}. Fig. \ref{fig:appendix:lamdiff} (red) compares values of \lam with and without simultaneous fitting of the sky spectrum. As before, the majority of cases show little change in \lam but there are a few galaxies where \lam changes significantly; a robust estimation of the standard deviation gives $\sigma_{\lambda-\lambda^{\prime}}=0.011$, which is less than the previous comparison and suggests that simultaneously fitting the sky is reducing the systematic error in \lam more than masking is. For completeness, we compare \lam calculated with masked spectra and with simultaneous sky fitting in Fig. \ref{fig:appendix:lamdiff} (green); a robust estimation of the standard deviation in this case gives $\sigma_{\lambda-\lambda^{\prime}}=0.016$, which further suggests that the masking technique is not as robust as the sky fitting technique. 

There are four galaxies in Fig. \ref{fig:appendix:lamdiff} for which \lam is very different when calculated using different techniques. Both the masking and sky fitting techniques agree on \lam for GMP3329 and GMP3352; indeed, the spectral fits when using masking or sky fitting techniques are much less affected by the sky. However, the masking and sky fitting techniques disagree on \lam for two galaxies: GMP3367 and GMP3792. For GMP3367, both techniques find identical velocity maps but differ slightly regarding the dispersion; however, both agree that this galaxy is an undisputed fast rotator. For GMP3792, the masking technique shows correlated errors in \V and \s for certain bins, artificially inflating \lam; the sky fitting technique finds no rotation in the velocity map and provides much better fits to the spectra.  

It is clear from these investigations that simultaneously fitting the sky spectrum is the most robust approach to calculating \V, \s and \lam; it is therefore the approach we adopt.

\bibliographystyle{mn2e}
\bibliography{MASTERBIB}

\bsp

\label{lastpage}

\end{document}